\documentclass[twocolumn,prd,nofootinbib,superscriptaddress]{revtex4-2}
\usepackage{color}
\usepackage{amsmath}
\usepackage{mathrsfs}
\usepackage{graphicx}
\usepackage{booktabs}
\usepackage{tikz}
\usepackage{bm}
\newcommand\mnras{MNRAS}

\newcommand\apjl{ApJL}
\newcommand\apjs{ApJS}

\newcommand\SkipNeuralNets[1]{}
\newcommand\skipme[1]{}
\newcommand\mysub[1]{\subsubsection{#1}}

\newcommand\qmstateproduct[2]{\left\langle#1|#2\right\rangle}

\newcommand\Y[1]{{{}_{#1}Y}}
\newcommand\lnL{ \ln {\cal L}}

\newcommand\optional[1]{}
\newcommand\unit[1]{{\rm #1}}
\newcommand\editremark[1]{{\color{red}#1}}

\newcommand{\jxWallLo}{10}
\newcommand{\jxWallHi}{25}

\newcommand{\tgSnrLo}{200}
\newcommand{\tgSnrHi}{1000}

\newcommand{\tgAreaHiMilliSq}{0.8}

\newcommand{\fvAgreePct}{7}

\newcommand{\fvSrate}{8192}
\newcommand{\fvFmax}{1024}

\newcommand{\jsrSnr}{600}
\newcommand{\jsrFmaxHz}{1024}

\newcommand{\jsrRateHz}{32768}
\newcommand{\jsrCredibleInnerPct}{50}
\newcommand{\jsrCredibleOuterPct}{90}
\newcommand{\jsrFisherAreaPhaseZero}{0.007641}
\newcommand{\jsrFisherAreaPhaseHalf}{0.007638}
\newcommand{\jsrFisherAreaPhaseDiffPct}{0.037}
\newcommand{\jsrSkyCovPhaseDiffPct}{0.09}

\newcommand{\jsrSimpsonRateHz}{8192}
\newcommand{\jsrSimpsonPhaseLnLDiff}{625.4}

\newcommand{\madCovDiffPct}{0.0046}

\newcommand{\jsrPortfolioNutsDiffPct}{0.1}
\newcommand{\jsrPortfolioFisherDiffPct}{3.2}

\newcommand{\qdRiRho}{652}
\newcommand{\qdRiNPhi}{8}
\newcommand{\qdRiNPsi}{8}
\newcommand{\qdRiGridOverInj}{777}
\newcommand{\qdRiGridOverRef}{901}
\newcommand{\qdRiOffsetGridDeg}{0.53}
\newcommand{\qdRiOffsetRatio}{17}

\newcommand{\vaLAngCostRatioHi}{0.063}
\newcommand{\vaLAngCostRatioLo}{0.947}

\newcommand{\vaLAngErrASpan}{256}

\newcommand{\vaLAngErrNRungs}{5}

\newcommand{\vaLAngErrTimesA}{0.0657}

\newcommand{\vaLAngErrTimesADef}{\rho^2/2}
\newcommand{\vaLAngErrTimesAScatter}{0.17}

\newcommand{\toaAccuracyBudget}{0.02}
\newcommand{\toaCandidateWorstLnL}{0.0035}
\newcommand{\toaCandidateWorstLnZ}{0.0012}

\newcommand{\toaNativeQRateHz}{4096}

\newcommand{\toaPregridFactor}{8}
\newcommand{\toaPregridRateHz}{32768}
\newcommand{\toaPregridSpacingMicroseconds}{30.5}

\newcommand{\toaQBandMaxHz}{1700}

\newcommand{\toaQNativeNyquistHz}{2048}
\newcommand{\toaValidationSnrMax}{652.31}

\newcommand{\ftGwCondAgreePct}{0.0}
\newcommand{\ftGwMargAgreePct}{2}
\newcommand{\ftGwShippedAreaPct}{141}
\newcommand{\ftGridConvergencePct}{0.20}
\newcommand{\ftRotationAreaPct}{1.6}
\newcommand{\ftFiniteArmAreaPct}{0.6}

\newcommand{\ftFsBBHthirtyaLIGO}{440}

\newcommand{\ftAreaEventBCEETK}{3300}

\newcommand{\ftFsNetEventBCEETK}{610}
\newcommand{\ftLnDmargEventBCEETK}{3.5}

\newcommand{\ftPsiCondEventBCEETK}{0.48}

\newcommand{\ftEventBTfixRatio}{0.25}
\newcommand{\ftEventBSubsampleBudgetNs}{73}
\newcommand{\ftEventBSigmaTUsAtThousand}{1.6}
\newcommand{\ftEventBFsKHzAtThousand}{610}
\newcommand{\ftBudgetFactor}{0.045}
\newcommand{\ftLogDistanceRange}{9.2}
\newcommand{\ftRotationStartHz}{5}
\newcommand{\ftEventBFminHz}{50}
\newcommand{\ftEventBFmaxHz}{1024}
\newcommand{\ftTwoGFminHz}{20}
\newcommand{\ftFmaxHz}{2048}

\newcommand{\ftGwfishVersion}{1.0.0}

\newcommand{\ftEventBTmargOverBlk}{4.0}
\newcommand{\ftPaperTfixRatioMeasured}{0.29}
\newcommand{\ftPaperAreaBandRhoSq}{2800}
\newcommand{\ftPaperAreaFixedRhoSq}{800}
\newcommand{\ftEventBAreaVsPaperBandPct}{20}
\newcommand{\ftLadderSigF}{66}
\newcommand{\ftLadderTblkMs}{2.4}

\newcommand{\ftLadderDeltaTUs}{240}
\newcommand{\ftLadderTkDerived}{0.1007}
\newcommand{\ftLadderTkFitted}{0.1002}

\newcommand{\ftLadderRungAgreePct}{2}
\newcommand{\ftLadderPsiCond}{0.59}
\newcommand{\ftLadderRhoLo}{41}
\newcommand{\ftLadderRhoHi}{652}

\newcommand{\ftLadderSigmaTHiUs}{3.7}
\newcommand{\ftLadderRhoMid}{163}
\newcommand{\ftLadderDistLo}{634}
\newcommand{\ftLadderDistHi}{39.6}
\newcommand{\ftLadderHsigThreshold}{2.5}
\newcommand{\ftLadderNdLo}{260}
\newcommand{\ftLadderNdMid}{4100}
\newcommand{\ftLadderNdHi}{6.6\times10^{4}}
\newcommand{\ftLadderHsigLo}{2.5}

\newcommand{\ftLadderHsigMid}{40}

\newcommand{\ftLadderHsigHi}{650}

\newcommand{\ftLadderHsigRefineA}{10}
\newcommand{\ftLadderErrRefineA}{0.33}
\newcommand{\ftLadderHsigRefineB}{2.5}
\newcommand{\ftLadderErrRefineB}{4.8\times10^{-4}}

\newcommand{\mpLReserveRungs}{41,82}
\newcommand{\mpLLocalRungs}{163,326}
\newcommand{\mpLMaxSelectedError}{7.6\times10^{-5}}
\newcommand{\mpLLocalCost}{0.15}

\newcommand{\aapAllAccept}{23}
\newcommand{\aapAllRows}{30}
\newcommand{\aapCostHi}{9.2}
\newcommand{\aapCostLo}{5.6}
\newcommand{\aapErrHi}{1.9\times10^{-3}}
\newcommand{\aapErrLo}{7.9\times10^{-5}}
\newcommand{\aapMemHi}{304}
\newcommand{\aapMemLo}{145}
\newcommand{\aapProdAccept}{9}
\newcommand{\aapProdRows}{10}
\newcommand{\aapReserveHi}{20.4}
\newcommand{\aapReserveLo}{0.37}
\newcommand{\aapRungs}{2}
\newcommand{\aapSeeds}{2}

\newcommand{\crtGPUJitSeconds}{18.59}
\newcommand{\crtGPUWarmSpeedup}{7.0}

\newcommand{\jaxEEMaxJS}{0.0043}

\newcommand{\jaxEERepeatLnLWithinThreeSigmaPct}{98.84}

\newcommand{\jaxEEIntrinsicPoints}{5000}

\newcommand{\bnsMOne}{1.6}
\newcommand{\bnsMTwo}{1.25}
\newcommand{\bnsDist}{100}
\newcommand{\bnsSnr}{23.8}
\newcommand{\bnsFmin}{30}

\newcommand{\bnsMc}{1.2293}
\newcommand{\bnsEta}{0.2462}
\newcommand{\bnsNile}{587}
\newcommand{\bnsNiter}{9}
\newcommand{\bnsNpts}{32}
\newcommand{\bnsNdraw}{1{,}600}
\newcommand{\bnsNext}{4{,}800}
\newcommand{\bnsNpooled}{320{,}000}
\newcommand{\bnsSkyArea}{17}

\newcommand{\bnsWorstBias}{0.6}
\newcommand{\bnsNInside}{6}
\newcommand{\bnsNparam}{6}

\newcommand{\bnsGpNfit}{576}

\newcommand{\bnsGpNsat}{2}
\newcommand{\bnsGpNsatRelaxed}{0}
\newcommand{\bnsGpHoldoutDefault}{2.3}
\newcommand{\bnsGpHoldoutRelaxed}{1.2}
\newcommand{\bnsGpHoldoutRelaxedLs}{1.2}
\newcommand{\bnsGpWidthGainMc}{40}
\newcommand{\bnsGpWidthGainEta}{45}
\newcommand{\bnsGpSigmaFDefault}{3.2}
\newcommand{\bnsGpSigmaFRelaxed}{93}
\newcommand{\bnsGpSigmaNDefault}{1.0}
\newcommand{\bnsGpSigmaNRelaxed}{0.44}
\newcommand{\bnsGpPeakDefault}{+0.92}
\newcommand{\bnsGpPeakRelaxed}{-0.19}
\newcommand{\bnsGpAccuracyGain}{1.9}

\newcommand{\bnsGpNrep}{3}

\newcommand{\bnsThreeSnr}{40}
\newcommand{\bnsThreeMOne}{1.6}
\newcommand{\bnsThreeMTwo}{1.25}
\newcommand{\bnsThreeFmin}{5}
\newcommand{\bnsThreeFmax}{512}
\newcommand{\bnsThreeRate}{1024}
\newcommand{\bnsThreeTerminal}{1600}

\newcommand{\bnsThreePool}{160{,}000}

\newcommand{\bnsThreeWarmUs}{35.4}
\newcommand{\bnsThreePrecomputeS}{302}
\newcommand{\bnsThreeFirstJaxS}{29}
\newcommand{\bnsThreeWorkerS}{375}

\newcommand{\bnsThreeCrediblePct}{90}
\newcommand{\bnsThreeInnerContourPct}{50}
\newcommand{\bnsThreeFisherFitPoints}{91}
\newcommand{\bnsThreeFisherFitRms}{0.11}
\newcommand{\bnsThreeFisherCvRms}{0.12}

\newcommand{\bnsThreeFisherMomentKl}{0.003}

\newcommand{\bnsThreeRefitKl}{0.025}
\newcommand{\bnsThreeRefitMaxQuantileShiftPct}{4.1}
\newcommand{\bnsThreeRefitMaxWidthShiftPct}{5.4}
\newcommand{\bnsThreeRefitHoldoutRms}{0.59}
\newcommand{\bnsThreeLowSkyLobePct}{14.7}
\newcommand{\bnsThreeLowSkyHalfOnePct}{14.8}
\newcommand{\bnsThreeLowSkyHalfTwoPct}{14.6}
\newcommand{\bnsThreeFirstHullPct}{100.0}
\newcommand{\bnsThreeRefitHullPct}{100.0}
\newcommand{\bnsThreePriorMcOuterPct}{0.0}
\newcommand{\bnsThreePriorEtaOuterPct}{0.0}

\newcommand\response[1]{{#1}}
\newcommand\mc{{{\cal M}_c}}
\def\RIT{Center for Computational Relativity and Gravitation, Rochester Institute of Technology, Rochester, New York
  14623, USA}

\def\UTAustin{University of Texas, Austin, TX 78712, USA}

\begin{document}
\title{Scaling RIFT 1: Extending RIFT to third-generation gravitational-wave analyses}
\author{R. O'Shaughnessy}
\affiliation{\RIT}
\author{A. Jan}
\affiliation{\UTAustin}
\author{J. Lange}
\affiliation{Cardiff University}
\author{R. Mechum}
\affiliation{\RIT}
\begin{abstract}
Interpretation of compact binary signals observed by third-generation gravitational-wave detectors is complicated by
their high signal-to-noise ratios, often reaching hundreds to thousands, and by time- and frequency-dependent detector
responses arising from Earth's rotation and the detectors' large physical extent.
Building on prior work for LISA, we implement generic, robust extensions of the RIFT parameter estimation pipeline that
enable interpretation of these sources using any existing waveform model, without recourse to intermediate pretrained data products
like reduced-order approximations or normalizing flows.
We demonstrate end-to-end  recovery of finite-size
binary-neutron-star injections in a three-site next-generation network to
$\mathrm{SNR}=\tgSnrHi$, localizing the source's extrinsic parameters to $\tgAreaHiMilliSq\times10^{-3}\,\mathrm{deg}^2$
in minutes on a single GPU.
\end{abstract}
\maketitle
\section{Introduction}
\label{sec:intro}
Ground-based gravitational-wave (GW) detectors---Advanced LIGO
\cite{2015CQGra..32g4001L}, Virgo
\cite{gw-detectors-Virgo-original-preferred,2015CQGra..32b4001A}, and KAGRA
\cite{2021PTEP.2021eA101A}---continue to identify coalescing compact binaries
\cite{LIGO-O3-O3b-catalog,LIGO-O3-O3a_final-catalog,LVK-GWTC4-observations,LVK-GWTC5-observations}, whose properties are characterized through
Bayesian inference
\cite{gw-astro-PE-lalinference-v1,gwastro-pe-bilby-2018,gwastro-PENR-RIFT,RevModPhys.94.025001}.
Third-generation (3G) observatories such as Cosmic Explorer and the Einstein Telescope
\cite{Evans2021CosmicExplorer,ET-Science-Blue-Book-2025} will change this problem in
several coupled ways.  Binary-neutron-star signals will remain in band for minutes to hours, long
enough for Earth's rotation to modulate the antenna response and arrival time
\cite{Whelan2024earthrotation}.  The loudest events will reach network signal-to-noise ratios
(SNRs) of hundreds or thousands, producing posterior widths that shrink approximately as
\(1/\mathrm{SNR}\).  Long detector arms also require a frequency-dependent response in the
sensitive band \cite{2025CQGra..42u5007B,2025PhRvD.112j2004B}.
Most preparation for this regime reduces the cost of individual likelihood evaluations.  Relative
binning and heterodyned likelihoods exploit the smooth ratio between a trial waveform and a
fiducial waveform
\cite{2018arXiv180608792Z,2021PhRvD.104j4054C}; multibanding adapts frequency resolution to the
chirp \cite{2021PhRvD.104d4062M}; and reduced-order quadrature projects the integral onto a
precomputed basis
\cite{gw-astro-ReducedOrderQuadraturePE-TiglioEtAl2014,gwastro-pe-ROM-IMRPv2-2016,2020PhRvD.102j4020M,2026arXiv260109819N}.
These approaches now support precession, higher modes, long 3G bandwidths, and in some cases
time-domain likelihoods
\cite{2024PhRvD.110h4085N,2026PhRvD.114d4007S}.  Reduced-order methods have enabled inference for
signals lasting roughly an hour or more
\cite{2021PhRvL.127h1102S,2026arXiv260614197G}.  Their basis construction can nevertheless become
the limiting step when Earth rotation and the finite-arm response are both retained at low
frequencies, precisely where much of the localization information accumulates
\cite{2025PhRvD.112j2004B}.
A complementary strategy amortizes inference by training a normalizing flow to map data directly
to posterior samples \cite{2021PhRvL.127x1103D}.  Such models can return complete
binary-neutron-star analyses in seconds \cite{2025Natur.639...49D} and have been applied to
high-redshift and hours-long 3G signals
\cite{2025PhRvD.112j3015S,2025ApJ...987L..17H}.  Their validity is tied to the prior, waveform
family, and noise distribution represented in training.  An alternative avoids a pretrained
inference model by combining a heterodyned likelihood, differentiable waveforms, and a
flow-enhanced gradient sampler
\cite{2023ApJ...958..129W,2024PhRvD.110h3033W}, but requires the waveform implementation itself to
be differentiable.
Other 3G complications include overlapping signals
\cite{2021PhRvD.104d4003S,2021MNRAS.507.5069A,2023MNRAS.523.1699J} and waveform errors that become
important at high SNR \cite{2020PhRvR...2b3151P}; we do not address either here.  We instead focus
on a limitation internal to the likelihood calculation.  Marginalization over reference phase,
polarization, arrival time, and distance uses numerical rules whose traditional grids are fixed by
the data or configuration, while the integrands narrow as \(1/\rho\).  A rule adequate for current
detections can therefore become biased at 3G amplitudes without an obvious failure signal.  The
same narrow, multimodal geometry also challenges samplers over the remaining extrinsic
coordinates.
RIFT is well suited to separate these problems.  Its factored likelihood uses waveform-model
outputs directly and precomputes the waveform-dependent inner products needed to evaluate many
extrinsic configurations
\cite{gwastro-PE-AlternativeArchitectures,gwastro-PENR-RIFT,gwastro-PENR-RIFT-GPU,gwastro-RIFT-Update,gwastro-RIFT_FinerNet}.
It therefore does not require a reduced basis, a waveform surrogate, or a differentiable
reimplementation of every waveform family.  The same architecture has recently been adapted to
LISA data \cite{gwastro-RIFT_LISA-Jan2024}.  For 3G observations, however, its detector response,
direct marginalizations, and extrinsic samplers must all remain accurate as the signal duration
and amplitude increase.
In this paper, we extend those three parts of the RIFT likelihood.  First, we generalize the
factored response to include Earth rotation and the finite-size, frequency-dependent response of
long arms while confining long-duration timeseries operations to a one-time precomputation.
Second, we derive amplitude-dependent resolution requirements for direct marginalization and
compare fixed-grid, structure-based, and peak-local rules for controlling their error and cost.
Third, we implement the fused extrinsic likelihood as an automatically differentiable JAX
function, use it to test gradient and normalizing-flow samplers, and introduce an adaptive
sequential-Monte-Carlo cloud for the narrow multimodal regime.  We validate the likelihood against
the production implementation and combine these developments in finite-size binary-neutron-star
injections in a three-site next-generation network at SNRs up to \(\tgSnrHi\).
This paper is one of four companion papers describing improvements developed along the
\texttt{rift\_O4c/d} line.  The core paper \cite{gwastro-RIFT-roboto-core} reviews the algorithm and
describes O4-era operating points, calibration marginalization, and cross-code likelihood tests.
The other companions address exported distance-resolved likelihood products
\cite{gwastro-RIFT-roboto-dgrid} and the hierarchical-inference ``hyperpipeline''
\cite{gwastro-RIFT-roboto-hyperpipe}.
This paper is organized as follows.  Section~\ref{sec:review} reviews the RIFT likelihood elements
needed here.  Section~\ref{sec:response} develops the time- and frequency-dependent detector
response.  Section~\ref{sec:jax_ile} describes the differentiable extrinsic samplers.
Section~\ref{sec:results_core_3g} presents validation and injection--recovery demonstrations, and
Section~\ref{sec:conclude} summarizes the results.  Appendices~\ref{ap:slowrot:modulation} and
\ref{ap:fd_precompute} give response-bookkeeping and implementation details; Appendix
\ref{ap:response-order} describes how to efficiently choose key operating point parameters for our detector response model;
Appendix~\ref{sec:quadrature} describes and Appendix~\ref{ap:quadrature} derives the direct-marginalization schemes.
\section{RIFT essentials}
\label{sec:review}
A coalescing compact binary can be completely characterized by its intrinsic
and extrinsic parameters.   The intrinsic parameters are necessary to characterize a binary's orbital inspiral and
merger trajectory, up to
spacetime symmetries.  For a quasicircular binary, this trajectory is specified by the binary's  (detector-frame) masses $m_{i,z}$, spins, and any quantities
characterizing matter in the system.  For an eccentric binary, these parameters must be supplemented by the orbital
eccentricity and mean anomaly.
Conversely, the extrinsic parameters are seven quantities reflecting flat spacetime symmetries: seven numbers needed to characterize its spacetime location and orientation.
We express masses in solar mass units and
 dimensionless spins in terms of Cartesian components $\chi_{i,x},\chi_{i,y}, \chi_{i,z}$, expressed
relative to a frame with $\hat{\mathbf{z}}=\hat{\mathbf{L}}$ and (for simplicity) at the orbital frequency corresponding to the earliest
time of computational interest (e.g., an orbital frequency of $\simeq 10 \unit{Hz}$).  We will use $\lambda,\theta$ to
refer to intrinsic and extrinsic parameters, respectively.
At a high level of abstraction, RIFT can be understood as a two-stage iterative process to interpret gravitational wave observations $d$ via comparison to
predicted gravitational wave signals $h(\bm{\lambda}, \bm\theta)$.   In one stage,
RIFT computes a marginal likelihood
\begin{equation}
 {\cal L}\response{({\bm \lambda})}\equiv\int  {\cal L}_{\rm full}(\bm{\lambda} ,\bm\theta )p(\bm\theta )d\bm\theta
\end{equation}
from the likelihood ${\cal L}_{\rm full}(\bm{\lambda} ,\theta ) $ of the gravitational wave signal in the multi-detector network,
accounting for detector response; see  \cite{gwastro-PE-AlternativeArchitectures,gwastro-PENR-RIFT} for a more detailed
specification.  This calculation is performed in parallel on a large number of candidate intrinsic parameters ${\bm
  \lambda}_\alpha$ by the integrate-likelihood-extrinsic (ILE) stage, which integrates the GW likelihood over extrinsic parameters.
In the second stage, RIFT interpolates and integrates.   Specifically, first it interpolates the likelihood information
accumulated from all previous iterations, making an approximation to ${\cal L}(\lambda)$ based on its
accumulated archived knowledge of marginal likelihood evaluations
$(\lambda_\alpha,{\cal L}_\alpha)$.  Second, using this approximation, it deduces the (detector-frame) posterior distribution
\begin{equation}
\label{eq:post}
p_{\rm post}=\frac{{\cal L}(\bm{\lambda} )p(\bm{\lambda})}{\int d\bm{\lambda} {\cal L}(\bm{\lambda} ) p(\bm{\lambda} )}.
\end{equation}
where prior $p(\bm{\lambda})$ is the prior on intrinsic parameters like mass and spin.
At the end of the iterative calculation, RIFT performs one final pass of ILE over posterior intrinsic draws $\lambda_k$,
fairly drawing some (fixed) number of extrinsic parameters $\theta_{k,p}$ from the  underlying ILE Monte
Carlo integral weighted samples.   These combined samples ($\lambda_k,\theta_{k,p}$) provide a full posterior for all
intrinsic and extrinsic parameters.   Postprocessing each sample provides derived parameters, including the source-frame masses $m^{\rm src}_{i}$.
This condensed review retains only the ingredients the present paper builds on: Section
\ref{sec:sub:L} describes how RIFT evaluates the gravitational wave likelihood
${\cal L}_{\rm full}$, Section \ref{sec:sub:Lmarg} how ILE performs the Monte Carlo
marginalization over extrinsic parameters, and Section \ref{sec:sub:CIP}, in brief, how the
marginal likelihood is interpolated into a posterior.  For the complete review --- including
iterative operation and coordinate choices, PP-test validation infrastructure, waveform and
calibration-marginalization interfaces, and LISA-RIFT --- see
\cite{gwastro-RIFT-Update,gwastro-RIFT_FinerNet} and the companion paper
\cite{gwastro-RIFT-roboto-core}.
\subsection{Review of the RIFT likelihood}
\label{sec:sub:L}
As described in previous work \cite{gwastro-PE-AlternativeArchitectures,gwastro-RIFT-Update}, RIFT uses physical
insight to decompose the overall inference calculation  into two stages: inference and marginal likelihoods for a fixed
physical binary, and inference about different physical binaries.
At a high level, RIFT expresses gravitational wave signals $h(t)$ in terms of physical basis signals $h_{lm}(t)$, associated
with a (spin-weighted) spherical harmonic decomposition of radiation in all possible emission directions.  This
decomposition allows RIFT to compute  cross-correlations between this basis and each detector's data; the likelihood for
arbitrary source orientations,  sky positions, and distances follows by a weighted average of these cross-correlation timeseries.
Using notation established in previous studies
\cite{gwastro-PENR-RIFT,gwastro-PENR-RIFT-GPU,gwastro-PE-AlternativeArchitectures}, the RIFT likelihood is expressed in
terms of a (spin-weighted) spherical harmonic decomposition of the complex gravitational wave strain
\begin{align} \label{eq:strain}
h(t,\vartheta,\phi;\bm{\lambda}) =  h_+(t,\vartheta,\phi;\bm{\lambda}) -
                                i h_\times (t,\vartheta,\phi;\bm{\lambda}) \, ,
\end{align}
Customarily its real and imaginary parts are denoted its two fundamental polarizations $h_+$ and $h_\times$.
Here, $t$ denotes time, $\vartheta$ and $\phi$ are the polar and azimuthal angles
for the direction of gravitational wave propagation away from the source.
The complex gravitational-wave strain can be written in terms of
spin-weighted spherical harmonics $\Y{-2}_{\ell m} \left(\vartheta, \phi \right)$ as
\begin{align} \label{eq:strain_mode}
h(t,\vartheta,\phi;\bm{\lambda}) =
\sum_{\ell=2}^{\infty} \sum_{m=-\ell}^{\ell} \frac{D_{\rm ref}}{D} h^{\ell m}(t;\bm{\lambda}) \Y{-2}_{\ell m} \left(\vartheta, \phi \right) \, ,
\end{align}
where the sum includes all available harmonic modes $h^{\ell m}(t;\pmb{\bm{\lambda}})$ made available by the model;  where
$D_{\rm ref}$ is a fiducial reference distance; and where $D$, the luminosity distance to the  source, is one of the
extrinsic parameters.
Each detector has (assumed constant) response functions $F_{+}$ and $F_\times$, such that the time-dependent strain
response has the form  $h_k(t) =F_{+,k} h_+(t_k) +
  F_{\times,k}h_\times(t_k)$ for the detector response $h_k$,
where $t_k=t_c - \vec{x}_k \cdot \hat{n}$ is the arrival time at the $k$th detector (at position $\vec{x}_k$)
for a plane wave propagating along $\hat{n}$ \cite{gwastro-PE-AlternativeArchitectures}.
We then substitute these expressions for $h_k$ into the standard Gaussian log-likelihood for stationary detector
noise \cite{gwastro-PE-AlternativeArchitectures},
thereby generating~\cite{gwastro-PE-AlternativeArchitectures}
\begin{widetext}
\begin{align}
\ln {\cal L}(\bm{\lambda}, \theta)
&= (D_{\rm ref}/D) \text{Re} \sum_k \sum_{\ell m}(F_k \Y{-2}_{\ell m})^* Q_{k,lm}(\bm{\lambda},t_k)\nonumber \\
&   -\frac{(D_{\rm ref}/D)^2}{4}\sum_k \sum_{\ell m \ell' m'}
\left[
{
|F_k|^2 [\Y{-2}_{\ell m}]^*\Y{-2}_{\ell'm'} U_{k,\ell m,\ell' m'}(\bm{\lambda})
}
 {
+  \text{Re} \left( F_k^2 \Y{-2}_{\ell m} \Y{-2}_{\ell'm'} V_{k,\ell m,\ell'm'} \right)
}
\right]
\label{eq:def:lnL:Decomposed}
\end{align}
\end{widetext}
where $F_k = F_{+,k} - i F_{\times,k}$ are the
complex-valued detector
response functions of the $k$th detector \cite{gwastro-PE-AlternativeArchitectures} and
the quantities $Q,U,V$ depend on $h$ and the data as
\begin{subequations}
\label{eq:QUV}
\begin{align}
Q_{k,\ell m}(\bm{\lambda},t_k) &\equiv \qmstateproduct{h_{\ell m}(\bm{\lambda},t_k)}{d}_k \nonumber\\
&= 2 \int_{|f|>f_{\rm low}} \frac{df}{S_{n,k}(|f|)} e^{2\pi i f t_k} \tilde{h}_{\ell m}^*(\bm{\lambda};f) \tilde{d}(f)\ , \\
{ U_{k,\ell m,\ell' m'}(\bm{\lambda})}& = \qmstateproduct{h_{\ell m}}{h_{\ell'm'}}_k\ , \\
V_{k,\ell m,\ell' m'}(\bm{\lambda})& = \qmstateproduct{h_{\ell m}^*}{h_{\ell'm'}}_k  \ .
\end{align}
\end{subequations}
The likelihood can be equivalently expressed as
\begin{align}
\ln {\cal L} &= \frac{D_{\rm ref}}{D} \text{Re}[ (F Y)^\dag Q]  \nonumber \\
 & - \frac{D_{\rm ref}^2}{4 D^2} [ (FY)^\dag U FY + (FY)^TV FY]
\label{eq:lnL:MatrixForm}
\end{align}
where this symbolic expression employs an implicit index-summation convention such that all naturally paired
  indices are contracted.  The result is an array of shape (time)$\times$(extrinsic).
For each candidate set of waveform parameters $\bm \lambda$, RIFT computes the inner product arrays $U,V$ and the inner
product timeseries $Q$.  Particularly for low-mass binaries, these precomputed quantities can be costly to evaluate, as
their calculation  involves manipulating long timeseries.  By contrast, the reduced quantities $Q$ and $\ln {\cal L}$
only need to be evaluated over a very short range of times, allowing for massive reduction in computational cost; see
\cite{gwastro-PENR-RIFT-GPU} for a detailed discussion.
In this work, we adopt significant improvements in how timeseries are manipulated and integrated, because
for the high-amplitude sources of interest to this study, the arrival times can often be constrained to lie well within
the timing resolution afforded by the underlying input data, as timing uncertainty scales as $1/\rho$ for $\rho$ the signal-to-noise ratio, with a coefficient set by the template's noise-weighted bandwidth (Appendix~\ref{ap:fisher_coeff}).  Every extrinsic evaluation requires high-accuracy interpolation (and integration) of
$Q_{k,\ell m}$.  In this study, we decouple the sampling rates of the timeseries used for the data $d$,  for
internally representing $Q$, and for performing the necessary time-marginalization.
Additionally, to increase efficiency, in this work we also explicitly marginalize in phase and polarization, using an
adaptive scheme that takes advantage of the restricted number of angular and polarization Fourier series modes needed
to characterize the log-likelihood.  Finally, to maximize efficiency across the full dynamic range of signals used in
this work, we swap from explicit integration over the whole domain to a fast, low-cost integral over locally Gaussian
approximations.  Appendix~\ref{ap:quadrature} describes our explicit time/phase/polarization marginalization schemes in more detail.
\subsection{The differentiable likelihood}
\label{sec:sub:jaxL}
For fixed intrinsic parameters $\bm{\lambda}$, the extrinsic likelihood of the preceding
subsection is a closed-form function of
$\bm\theta=(\alpha,\delta,\psi,\iota,\phi_{\rm ref},D)$.  Once the inner-product arrays
$U_{k,\ell m,\ell'm'}$ and $V_{k,\ell m,\ell'm'}$ and the cross-correlation timeseries
$Q_{k,\ell m}(t)=\qmstateproduct{h_{\ell m}(t)}{d}_k$ have been precomputed,
Eq.~\eqref{eq:lnL:MatrixForm} requires only the antenna pattern
$F_k(\alpha,\delta,\psi)$, arrival-time delay $t_k$, spin-weighted harmonics
$\Y{-2}_{\ell m}(\iota,\phi_{\rm ref})$, and distance scaling $D_{\rm ref}/D$.  We implement this
fused reduction in JAX, making $\ln\mathcal{L}(\bm{\lambda},\bm\theta)$ automatically
differentiable and compatible with \texttt{jit} and \texttt{vmap}.  Frame reading, power-spectral-density
handling, and the precomputation of $Q$, $U$, and $V$ remain unchanged from the production code
\cite{gwastro-PENR-RIFT-GPU}.  On identical inputs, the differentiable path reproduces the
production array-vector likelihood to $\sim\!10^{-13}$, its gradients match finite differences,
and its harmonics agree with \texttt{lal} for $\ell=2,\ldots,8$. A cross-method comparison
on publicly released H1--L1 strain from GW240925 also finds agreement.
The separate full-pipeline comparison below uses observed H1--L1 strain from GW240426.
Both are analyses of LVK detector data; the third-generation demonstrations use simulated signals in zero noise.
The extrinsic likelihood is sharply peaked in several directions.  Its derivatives provide local
Fisher information for sampler initialization and Laplace approximations, while gradient-based
samplers can follow the peak instead of locating it through adaptive Monte Carlo
(Eq.~\ref{eq:lnL:MonteCarlo}).  The resulting toolkit includes NUTS
\cite{Hoffman2014NUTS}, normalizing-flow Monte Carlo
\cite{pmlr-v37-rezende15,Gabrie2022flowMC,Wong2023flowMC}, Fisher preconditioning, and adaptive
sequential Monte Carlo \cite{Neal2001AIS,DelMoral2006SMC}, none of which was available to the
legacy adaptive-Cartesian integrator \cite{gwastro-RIFT-Update}.
That said, the JAX likelihood also performs well with the existing AV Monte Carlo sampler, because in fact the JAX
implementation is faster (by roughly $\crtGPUWarmSpeedup$ times) than our original (cupy) GPU implementation.
While a JAX implementation incurs some one-time startup cost (roughly $\crtGPUJitSeconds\,$s for this test,
associated with compilation cost), this cost is both modest and distributed over the many evaluations each ILE worker
typically performs.
When using a derivative-aware sampler with the differentiable likleihood, we always marginalize over several extrinsic parameters including distance via direct quadrature.
When marginalizing over distance, we choose the integration range (and method) to cover and resolve the narrow distance
peak in the likelihood.
More broadly, differentiability allows the quadratures to be sized from the integrand.
Automatic differentiation provides a gradient that locates the
maximum and a Fisher matrix $\Gamma$ whose inverse-covariance determines the range needed in each parameter (from the
covariance matrix $\Gamma^{-1}$).
This estimate can allow us to adaptively choose the numerical methods employed for each parameter when the source (and
thus $Q,U,V$) are specified.
Appendix~\ref{ap:quadrature} describes how the various marginalization schemes take advantage of information about the
characteristic scale of the integrand.
The driver \texttt{integrate\_likelihood\_extrinsic\_jax} adopts the same command-line arguments and input/output
formats,
so changing the ILE version does not change the surrounding workflow.
Some argument options remain unimplemented in JAX-ILE, including cross-process proposal-state persistence
(Appendix~\ref{ap:sampler_menu}) and some waveform families.
\subsection{Evaluating the marginalized likelihood}
\label{sec:sub:Lmarg}
Given the likelihood ${\cal L}_{\rm full}(\bm\lambda,\bm\theta)$, other versions of RIFT evaluate the marginal likelihood via an adaptive Monte Carlo
integrator:
\begin{align}
\label{eq:lnL:MonteCarlo}
{\cal L}(\bm\lambda) \simeq \frac{1}{N} \sum_k {\cal L}_{\rm full}(\bm\lambda,\theta_k) p(\bm\theta_k)/p_s(\bm\theta_k)
\end{align}
RIFT provides multiple  adaptive Monte Carlo techniques \cite{gwastro-RIFT-Update} to perform this and any other
integral.
Due to its ubiquitously robust performance, almost all conventional RIFT calculations currently use the adaptive volume (AV)
integrator \cite{gw-astro-mergers-VTiwariPE,gwastro-RIFT_FinerNet}.
Additionally, as described in the companion paper \cite{gwastro-RIFT-roboto-core},
the AV and other RIFT integrators have recently been made \emph{warm-startable} and \emph{reusable}.
The substantial cost of adapting a
proposal to the extrinsic likelihood need not be paid independently by every one of the many ILE evaluations that share similar
structure.  The AV live volume can be bootstrapped directly from prior information -- a previous run's fair-draw samples, a
Fisher matrix at the maximum-likelihood point, or a serialized live-volume state -- and the oracle and normalizing-flow proposals
can likewise be persisted and reloaded.  Whether a warm start can bias the result depends on how the seed is used.
A seed that only shapes the sampling distribution $p_s$ leaves the estimator of
Eq.~\eqref{eq:lnL:MonteCarlo} unchanged and cannot bias it, because the importance weights correct for whatever was
sampled.  A seed that initializes the adaptive integrator's live \emph{volume} is different: that integrator only ever
contracts, so likelihood outside the seeded region is never evaluated and cannot be recovered, and too tight a seed
therefore loses mass; see companion paper \cite{gwastro-RIFT-roboto-core} for more discussion and necessary safety
margins the code must adopt to avoid losing mass.
This matters most for the very high-amplitude sources that motivate this work: as the network amplitude $\rho$ grows, the
extrinsic posterior occupies a rapidly shrinking fraction ($\sim\rho^{-d}$ over $d$ extrinsic dimensions) of the prior volume,
and a Cartesian integrator started from the prior can fail to locate it within any practical evaluation budget;
a warm-started illustration on a loud real event appears in the companion paper \cite{gwastro-RIFT-roboto-core}.
\subsection{Likelihood interpolation and posterior distributions}
\label{sec:sub:CIP}
The construct-intrinsic-posterior (CIP) stage interpolates the marginal likelihood and draws from the intrinsic posterior.
To estimate ${\cal L}$ from discrete samples $\lambda_\alpha,{\cal L}_\alpha$, RIFT previously  deploys two unstructured
interpolation techniques:  Gaussian process regression and random forest regression.
Given the extremely high training and evaluation cost for Gaussian processes (as well as the finicky training necessary
to optimize their model hyperparameters), in practice we continue to almost exclusively
recommend using random forest regression for production work.
In \emph{this} work, however, we use Gaussian process regression for the controlled
binary-neutron-star recovery of Section~\ref{sec:results:bns}, whose smooth likelihood spans a
larger dynamic range than the default signal-amplitude bound can represent.   Appendix~\ref{ap:gp_scale}
describes improvements to the estimated Gaussian process kernel lengthscales, so the GP can adapt correctly
to sources with third-generation-scale posteriors in mass and amplitude.
Given the likelihood, fair samples from the posterior distribution are generated by the following two-step process,
described in the RIFT paper.
First, using the likelihood estimate $\hat{\cal L}_{\rm marg}$ and the same AV adaptive Monte Carlo integrator described above,
we perform the Monte Carlo integral $\int d{\bm \lambda} \hat{\cal L}_{\rm marg} p({\bm \lambda})$, producing sample
locations ${\bm \lambda}_k$ and associated weights $w_k= \hat{\cal L}_{\rm marg} p(\lambda)/p_s(\lambda)$.
Second, we make a fair draw from these weighted samples.
In typical practice, the Monte Carlo integration(s) used to generate the posterior are  performed in
parallel,  with each CIP worker instance terminating once a threshold $n_{\rm eff}$ is reached, where as described at length in
Wofford et al \cite{gwastro-RIFT-Update} RIFT uses an effective sample estimate
\begin{align}
\label{eq:neff}
n_{\rm eff} = \frac{\sum p_k}{\max \{p_k \} }
\end{align}
expressed for convenience in terms of normalized sample probabilities $p_k = w_k/\sum_q w_q$.
The value $1/n_{\rm eff}$ is the largest discontinuous jump in the estimator $\hat{P}(<x) = \sum_k p_k \theta(x_k-x)$
for any one-dimensional cumulative probability distribution $P(<x)$ derived from the full samples.
In this work, unless otherwise noted we use $n_{\rm eff}/N$ to characterize our sampling efficiency.
\section{A generalized detector response: Earth rotation and finite-size arms}
\label{sec:response}
RIFT adopts multiple simplifications that are well-adapted to current interferometers.  For
example, like most contemporary GW analysis strategies, it treats each detector's response as a
pair of constants, neglecting both its drift during the observation and its frequency dependence.
In the third-generation era, however, these approximations break down for typical binary neutron star sources, which will
in band for tens of minutes to hours, over which the Earth rotates appreciably, and
$10$--$40$-km arms acquire a frequency-dependent, finite-size response within the observing band.
In this section we generalize the factored likelihood of Section~\ref{sec:sub:L} to both effects, using two controlled
power-series expansions whose order can be chosen to achieve the necessary accuracy.
Throughout, we write the extrinsic parameters as
$\bm\theta=(\alpha,\delta,\psi,\iota,\phi_{\rm ref},D)$: right ascension, declination and
polarization angle, which fix the source direction $\hat n$ and the polarization basis; the
inclination and reference orbital phase, which fix the emission direction; and the luminosity
distance.
To establish notation and ground the subsequent derivation, we first review why the likelihood factors under the
assumptions previously adopted for RIFT: assuming the detectors are short (so the detector response functions are
time-independent, frequency-independent constants).
We first ecompose the signal in terms of angular modes as
\begin{equation}
h_Y(t;\bm\lambda,\iota,\phi_{\rm ref})\equiv\sum_{\ell m}\Y{-2}_{\ell m}(\iota,\phi_{\rm ref})\,
   h_{\ell m}(t;\bm\lambda),
\label{eq:resp:hY}
\end{equation}
and let $\mathcal{T}_\tau$ denote translation in time,
\begin{equation}
(\mathcal{T}_\tau g)(t)\equiv g(t-\tau),\qquad
\widetilde{\mathcal{T}_\tau g}(f)=e^{-2\pi i f\tau}\,\tilde g(f).
\label{eq:resp:translation}
\end{equation}
The strain recorded by the $k$th detector is then the real part of a single complex object, the
\emph{response-folded template} $\mathfrak{h}_k\equiv F_k\,\mathcal{T}_{t_k}h_Y$,
\begin{equation}
h_k(t)=\mathrm{Re}\big[F_k\,(\mathcal{T}_{t_k}h_Y)(t)\big],
\label{eq:resp:strain:static}
\end{equation}
with $F_k=F_{+,k}-iF_{\times,k}$ the complex antenna factor and $t_k$ the geometric arrival delay
of Section~\ref{sec:sub:L}.  Because $h_k$ is real and $d$ is real, the two inner products that
build the log-likelihood $\lnL=\qmstateproduct{h_k}{d}_k-\tfrac12\qmstateproduct{h_k}{h_k}_k$
follow from $\qmstateproduct{\mathrm{Re}[zg]}{d}=\mathrm{Re}[z^*\qmstateproduct{g}{d}]$ for any
complex constant $z$:
\begin{subequations}
\label{eq:resp:static:ip}
\begin{align}
\qmstateproduct{h_k}{d}_k&=\mathrm{Re}\big[F_k^*\,
   \qmstateproduct{\mathcal{T}_{t_k}h_Y}{d}_k\big],\\
\qmstateproduct{h_k}{h_k}_k&=\tfrac12|F_k|^2\,
   \qmstateproduct{\mathcal{T}_{t_k}h_Y}{\mathcal{T}_{t_k}h_Y}_k \nonumber\\
&\quad+\tfrac12\mathrm{Re}\big[F_k^2\,
   \qmstateproduct{(\mathcal{T}_{t_k}h_Y)^*}{\mathcal{T}_{t_k}h_Y}_k\big].
\end{align}
\end{subequations}
Two properties then collapse Eq.~\eqref{eq:resp:static:ip} into
Eqs.~\eqref{eq:def:lnL:Decomposed}--\eqref{eq:lnL:MatrixForm}.  First, $F_k$ is a
\emph{constant}, so it passes through every inner product as a scalar prefactor and never touches
the frequency integral.  Second, $\mathcal{T}_\tau$ is \emph{unitary} for the noise-weighted
inner product: $\qmstateproduct{\mathcal{T}_\tau a}{\mathcal{T}_\tau b}=\qmstateproduct{a}{b}$,
and --- because complex conjugation reflects the frequency argument, so that the two translation
phases again cancel --- $\qmstateproduct{(\mathcal{T}_\tau a)^*}{\mathcal{T}_\tau b}
=\qmstateproduct{a^*}{b}$ as well.  The arrival time therefore cancels identically from the
quadratic term, so $U$ and $V$ of Eq.~\eqref{eq:QUV} may be evaluated once at a fiducial epoch,
while $t_k$ survives only as the time at which the precomputed timeseries
$Q_{k,\ell m}(t)=\qmstateproduct{\mathcal{T}_t h_{\ell m}}{d}_k$ is read.  Expanding $h_Y$ in the
harmonic basis then returns Eq.~\eqref{eq:def:lnL:Decomposed} line for line.  These two
properties are the whole content of ``precompute and marginalize'': every expensive object
depends only on $\bm\lambda$, and every extrinsic parameter enters through a scalar contraction.
Both of the two approximations adopted previously in RIFT break down for third-generation interferometers.   For
long-duration signals, a rotating Earth makes
the antenna factor and the propagation delay functions of time,
\begin{equation}
h_k(t)=\mathrm{Re}\big[F_k(t)\,h_Y\!\big(t-t_k(t)\big)\big],
\label{eq:resp:strain:rot}
\end{equation}
so that $F_k$ no longer factors out and the propagation is no longer any single translation
$\mathcal{T}_\tau$.
Long arms make the antenna factor a function
of frequency, $F_k\to F_k(f) $.
Nonetheless, we can refactor the relevant operator products, allowing us to recover the same structure of RIFT's
customary likleihood, albeit with a richer set of basis waveforms and overlap arrays.
Section~\ref{sec:slowrot}
carries this out for the time-dependent response and Section~\ref{sec:finitesize} for the
frequency-dependent one.  The two are independent and compose.
\subsection{Earth rotation: a slowly time-varying response}
\label{sec:slowrot}
The static-response approximation is excellent for the
$\mathcal{O}(1\text{--}10\,\mathrm{s})$ signals analyzed today, but it fails for \emph{long}
signals, over which the Earth rotates appreciably during the observation: binary neutron stars
already last minutes in current detectors, and the loudest binary neutron stars in
next-generation detectors last $\sim\!90$ minutes, over which the sidereal phase
$\Omega_\oplus T\approx0.4\,\mathrm{rad}$ ($\Omega_\oplus=7.29\times10^{-5}\,\mathrm{s}^{-1}$)
and both the antenna pattern and the light-travel delay drift.  The drift both biases parameter
estimation if ignored and carries information, since a single detector's rotation localizes a
long signal on its own.
\mysub{A two-stage expansion}
The time-dependent  detector strain  $\mathfrak{h}_{k}(t)=F_k(t)\,h_Y(t-t_k(t))$ carries time dependence in a prefactor
and in its (non-uniform) arrival time.   To preserve RIFT's likleihood structure and efficient precomputation, we seek to represent the strain in
terms of a sum of elementary basis functions derived from the modes $h_{\ell m}(t)$.
The antenna response functions
are quadratic forms in polarization tensors that rotate rigidly with the Earth, so $F_k(t)$ is
\emph{exactly} band-limited to five sidereal harmonics,
\begin{equation}
F_k(t)=\sum_{n=-2}^{2}A_n(\delta,\psi)\,e^{\,in g(t)},\qquad
g(t)=\mathrm{GMST}(t)-\alpha,
\label{eq:slowrot:Fharm}
\end{equation}
with no content beyond $|n|=2$~\cite{Whelan2024earthrotation}.  The five complex coefficients
$A_n$ are closed-form functions of the detector response tensor, the source declination, and the
polarization angle \emph{only} --- notably \emph{independent} of right ascension and of time,
which enter solely through the sidereal phase $g(t)$.  Since $\mathrm{GMST}$ advances uniformly
at $\Omega_\oplus$, Eq.~\eqref{eq:slowrot:Fharm} states that multiplication by $F_k(t)$ is a
five-term sum of \emph{modulation} operators
\begin{equation}
(\mathcal{M}_n g)(t)\equiv e^{\,in\Omega_\oplus(t-t_{\rm ref})}\,g(t),
\label{eq:slowrot:modop}
\end{equation}
each of which is sky-independent, with all of the sky, polarization and reference-time dependence
carried by the scalar prefactors $A_n\,e^{\,in g(t_{\rm ref})}$.
The delay factor can be restructured using the conventional representation of translation as an (infinite) sum over
derivatives.  [A time-varying delay is a
time-varying phase $e^{2\pi if\,t_k(t)}$ that entangles time and frequency, and it does not
reduce to a few slow scalars in the frequency domain.]
Specifically, we  expand it in the
\emph{time} domain, about a fixed reference delay $\tau_0$, in powers of the drift
$\delta t_k(t)=t_k(t)-\tau_0$:
\begin{equation}
h_{\ell m}\!\big(t-t_k(t)\big)=\sum_{p\ge0}\frac{\big(-\delta t_k(t)\big)^p}{p!}\,
   h_{\ell m}^{(p)}(t-\tau_0),
\label{eq:slowrot:delay}
\end{equation}
where $h^{(p)}_{\ell m}\equiv d^ph_{\ell m}/dt^p$.  This is the operator expansion of a
translation about $\tau_0$ in powers of its generator, $\mathcal{T}_{\tau_0+\delta}=
\mathcal{T}_{\tau_0}e^{-\delta\,\partial_t}$.
It differentiates
the \emph{full} inertial-frame mode timeseries, so whatever sidebands and mode mixing higher
multipoles and precession have imprinted on $h_{\ell m}(t)$ are carried exactly, with no per-mode
stationary-phase approximation and no time--frequency map.  And the drift is itself band-limited:
the detector position $\vec{x}_k(t)$ rotates rigidly, so $\delta t_k(t)$ carries only the
harmonics $n\in\{0,\pm1\}$, and $(-\delta t_k)^p$ therefore carries $|n|\le p$ --- so
multiplication by it is again a finite sum of the operators $\mathcal{M}_n$.
Composing the two stages leaves a single finite sum.  Collecting a delay-derivative order $p$ and
a sidereal-harmonic index $n$ into a composite index $a=(p,n)$, the response-folded template is
\begin{equation}
\begin{split}
\mathfrak{h}_{k,\ell m}(t)&=\sum_{a}C^{\,k}_a(\alpha,\delta,\psi;t_{\rm ref})\,
   \chi^{(a)}_{\ell m}(t),\\
\chi^{(p,n)}_{\ell m}&\equiv\mathcal{M}_n\,\mathcal{T}_{\tau_0}\,h^{(p)}_{\ell m},
\end{split}
\label{eq:slowrot:master}
\end{equation}
with the coefficients obtained by multiplying out the two harmonic lists,
$C^k_{(p,\tilde n)}=\tfrac1{p!}\sum_{n+m=\tilde n}A_n e^{\,ing(t_{\rm ref})}
[(-\delta t_k)^{\ast p}]_m$, where $\ast p$ denotes the $p$-fold discrete convolution of the
delay-drift harmonics; the sum over $n$ runs over $|n|\le2+p$.  Equation~\eqref{eq:slowrot:master}
is the key structural statement of this section.  The \emph{elementary modulated templates}
$\chi^{(a)}_{\ell m}$ are intrinsic-only and sky-independent, so their inner products with the
data and with each other are precomputed by exactly the existing machinery of
Eq.~\eqref{eq:QUV} applied to $\chi^{(a)}_{\ell m}$ in place of $h_{\ell m}$; \emph{all} sky,
polarization and reference-time dependence lives in the analytic scalars $C^k_a$.  Provided the
sum over $a$ is short, the sky remains an extrinsic parameter and the fast marginalization
survives, gaining only short inner sums.
\mysub{Reference-time bookkeeping}
The expansion runs on two clocks, and keeping them apart is the whole of its correctness.  The
response modulation is a function of \emph{absolute} time: the Earth's orientation, and hence
the antenna harmonics of Eq.~\eqref{eq:slowrot:Fharm} and the delay drift of
Eq.~\eqref{eq:slowrot:delay}, depend on when the wave arrives, not on where a template is placed.
The elementary templates, by contrast, are built once and slid in time by an FFT correlation, so
the bank holds them as functions of the template's own \emph{intrinsic} time --- that is, it
realizes the operator ordering $\mathcal{T}_{t_k}\mathcal{M}_n$, modulate first and then place,
whereas Eq.~\eqref{eq:resp:strain:rot} demands $\mathcal{M}_n\mathcal{T}_{t_k}$, place first and
then modulate in absolute time.  The two orderings differ by a c-number,
\begin{equation}
\mathcal{M}_n\,\mathcal{T}_\tau=e^{\,in\Omega_\oplus\tau}\;\mathcal{T}_\tau\,\mathcal{M}_n ,
\label{eq:slowrot:commutator}
\end{equation}
so the modulation and the propagation simply do not commute.  Referred to $t_{\rm ref}$, the
residual factor is the \emph{arrival-time post-phase}, and it belongs to the coefficient:
\begin{equation}
\tilde C^k_a\equiv C^k_a\,e^{\,in_a\Omega_\oplus(t_k-t_{\rm ref})}.
\label{eq:slowrot:postphase}
\end{equation}
Its argument is the geometric propagation delay, of order $10\,\mathrm{ms}$, and it vanishes only
if the template happens to be placed at the reference time itself.  One rule then makes the
scheme correct: $\tilde C^k_a$, and never $C^k_a$, is the coefficient of $\chi^{(a)}_{\ell m}$
--- in the data term built from $Q$ \emph{and} in the model norm built from $U$ and $V$.  Because
Eq.~\eqref{eq:slowrot:commutator} is a scalar identity the post-phase costs nothing to apply, but
applying it to one term and not the other evaluates $\qmstateproduct{d}{h}$ and
$\qmstateproduct{h}{h}$ for two \emph{different} templates.  Appendix~\ref{ap:slowrot:modulation}
records what that costs, why the tempting shortcut of moving the modulation onto the data instead
is invalid, and why the natural convergence and brute-force tests are blind to both errors.
\mysub{The likelihood with the sidereal index realized}
With Eq.~\eqref{eq:slowrot:master} and the post-phase in hand the generalized likelihood can be
written out explicitly.  Define the precomputed arrays exactly as in Eq.~\eqref{eq:QUV}, but on
the elementary modulated templates rather than the modes:
\begin{subequations}
\label{eq:slowrot:QUV}
\begin{align}
Q_{k,a\ell m}(\bm{\lambda},t) &\equiv
   \qmstateproduct{\mathcal{T}_t\,\chi^{(a)}_{\ell m}(\bm{\lambda})}{d}_k ,\\
U_{k,a\ell m,a'\ell'm'}(\bm{\lambda}) &\equiv
   \qmstateproduct{\chi^{(a)}_{\ell m}}{\chi^{(a')}_{\ell'm'}}_k ,\\
V_{k,a\ell m,a'\ell'm'}(\bm{\lambda}) &\equiv
   \qmstateproduct{[\chi^{(a)}_{\ell m}]^*}{\chi^{(a')}_{\ell'm'}}_k .
\end{align}
\end{subequations}
Substituting Eq.~\eqref{eq:slowrot:master} into Eq.~\eqref{eq:resp:static:ip} --- which is legal
because the $\tilde C^k_a$ are constants, the property Eq.~\eqref{eq:resp:strain:rot} had
destroyed and the expansion has now restored --- gives
\begin{widetext}
\begin{align}
\ln {\cal L}(\bm{\lambda}, \bm\theta)
&= (D_{\rm ref}/D)\,\text{Re} \sum_k \sum_{a}\sum_{\ell m}
   \big(\tilde C^k_a\,\Y{-2}_{\ell m}\big)^*\,Q_{k,a\ell m}(\bm{\lambda},t_k)\nonumber \\
&\quad -\frac{(D_{\rm ref}/D)^2}{4}\sum_k \sum_{a a'}\sum_{\ell m \ell' m'}
\Big[
(\tilde C^k_a)^*\tilde C^k_{a'}\,[\Y{-2}_{\ell m}]^*\Y{-2}_{\ell'm'}\,
   U_{k,a\ell m,a'\ell' m'}
\nonumber\\
&\hspace{6.5em}
+  \text{Re} \big( \tilde C^k_a\tilde C^k_{a'}\,\Y{-2}_{\ell m} \Y{-2}_{\ell'm'}\,
   V_{k,a\ell m,a'\ell'm'} \big)
\Big] .
\label{eq:slowrot:lnL:Decomposed}
\end{align}
\end{widetext}
This is Eq.~\eqref{eq:def:lnL:Decomposed} with one index added and $F_k$ replaced by
$\tilde C^k_a$.  The matrix form makes the point compactly: writing $\mathcal{C}_k$ for the
vector with components $\mathcal{C}_{k,(a,\ell m)}=\tilde C^k_a\,\Y{-2}_{\ell m}$ over the
composite index $A=(a,\ell m)$,
\begin{align}
\ln {\cal L} &= \frac{D_{\rm ref}}{D} \text{Re}\big[ \mathcal{C}^\dag Q\big]  \nonumber \\
 & \quad - \frac{D_{\rm ref}^2}{4 D^2}
   \big[ \mathcal{C}^\dag U \mathcal{C} + \mathcal{C}^T V \mathcal{C}\big] ,
\label{eq:slowrot:lnL:MatrixForm}
\end{align}
which is Eq.~\eqref{eq:lnL:MatrixForm} verbatim, with the harmonic index $\ell m$ widened to
$(a,\ell m)$ and the antenna factor $F_k$ widened to $\tilde C^k_a$.  Setting $\Omega_\oplus\to0$
collapses the index to $a=(0,0)$ with $\tilde C^k_{(0,0)}\to F_k$ and recovers the static
likelihood identically.
Three consequences of this form are worth stating.  The extrinsic marginalization is untouched:
$\mathcal{C}_k$ is an analytic function of $(\alpha,\delta,\psi,\iota,\phi_{\rm ref})$ and the
distance still enters as $D_{\rm ref}/D$, so the sky remains an extrinsic parameter and the
sampler of Section~\ref{sec:sub:Lmarg} sees a likelihood of the same shape.  The bank needs no
templates beyond $\{\chi^{(a)}_{\ell m}\}$: because
$[\chi^{(p,n)}_{\ell m}]^*=\mathcal{M}_{-n}\mathcal{T}_{\tau_0}[h^{(p)}_{\ell m}]^*$, the $V$
array of Eq.~\eqref{eq:slowrot:QUV} is the ordinary $V$-type overlap of the derivative modes with
the sidereal index reflected.  And the post-phase enters the quadratic term only through a
harmonic difference --- as $n_{a'}-n_a$ in the $U$ contraction, whose coefficient is
$(\tilde C^k_a)^*\tilde C^k_{a'}$, and identically in the $V$ contraction once the reflection is
taken into account, its coefficient being $\tilde C^k_{(p_a,-n_a)}\tilde C^k_{a'}$ --- so it
costs one phase per distinct difference rather than one per pair.  The precompute grows by a bounded factor --- the size of the bank, about $5\times$ for
the five antenna harmonics and a further factor $\sim(p_{\max}+1)$ for the delay derivatives ---
and the extrinsic stage grows only by the inner sums over $a$ in
Eq.~\eqref{eq:slowrot:lnL:Decomposed}.  Appendix~\ref{ap:fd_precompute} describes how the bank
is built without ever materializing a family of time-domain templates.
\mysub{Two limiting cases useful for validation tests}
The expansion of Eq.~\eqref{eq:slowrot:master} contains two nested approximations --- the
amplitude modulation and the delay drift --- and it is worth isolating them, because each has a
clean physical regime of validity.  In our internal code tests shipped as part of the RIFT release, we turn these two
factors (henceforth denoted as Paths A and B) on and off independently, providing controlled limits which can more easily be validated than the full response.
\emph{Path~A} (omitting the Earth's rotation from the delay time) keeps only $p=0$: the delay is held fixed at $\tau_0$ and only the antenna
amplitude drifts, so $a=(0,n)$ with $C^k_{(0,n)}=A_n e^{\,ing(t_{\rm ref})}$ and the bank is five
templates.  For the amplitude this is not an approximation at all --- Eq.~\eqref{eq:slowrot:Fharm}
terminates at $|n|=2$, so Path~A carries the antenna drift \emph{exactly}.  What it neglects is
the $p\ge1$ terms of Eq.~\eqref{eq:slowrot:delay}, whose size relative to the leading term is
$\delta t_k\,|\dot h_{\ell m}/h_{\ell m}|$.  The logarithmic derivative is set by the
instantaneous frequency, $|\dot h_{\ell m}/h_{\ell m}|\sim 2\pi f$, so the neglected fraction is
$\sim2\pi f\,\delta t_k$: Path~A is accurate whenever the signal's power sits at frequencies low
enough that the delay drift is a small fraction of a wave period.
\emph{Path~B} (using a truncated power series for the delay time) retains  those terms to order $p_{\max}$, with error of order
the first omitted term, $\sim(2\pi f\,\delta t_k)^{p_{\max}+1}/(p_{\max}+1)!$.  This expansion
converges rapidly once the expansion parameter is below unity and fails abruptly once it is not.
At $\Omega_\oplus\to0$, the two series are equivalent.
The worst realistic case, a $90$-minute binary neutron star in a
next-generation detector, needs only $p_{\max}\sim2$--$3$; often $p_{\max}\simeq0$ or $1$ suffices.
\subsection{Long arms: a frequency-dependent response}
\label{sec:finitesize}
The static-response assumption fails a second, independent way once the arms are long compared
with the gravitational wavelength.  Equation~\eqref{eq:resp:strain:static} treats each detector as
sampling the metric perturbation at a point.  In fact the observable is a light travel time: a
photon launched down an arm at time $t$ returns at $t+2T$, having sampled the metric continuously
along the way, so what the interferometer reports is a phase-weighted \emph{average} of $h$ over
the light-crossing time $T=L/c$ rather than its instantaneous value.  Averaging is exact only if
$h$ is constant over $T$; once $fT$ is not negligible the coefficient multiplying each
polarization becomes a \emph{filter} rather than a number.  The resulting high-frequency
corrections to the ground-based response are standard \cite{gwastro-GroundBasedResponse-Whelan2008},
and have recently been carried into parameter estimation directly, where neglecting them --- or
neglecting the Earth's rotation of Section~\ref{sec:slowrot} --- is shown to bias recovered
parameters \cite{2025CQGra..42u5007B}.  Our contribution is not the response itself but a
factorization of it that leaves RIFT's precompute-and-marginalize architecture intact.
Concretely, integrating the one-way phase accumulated by a photon travelling along $\hat a$
through a plane wave arriving from $\hat n$, and adding the return trip, gives the arm transfer
function
\begin{equation}
\begin{split}
\tilde D(\hat a,\hat n,f)=\frac{e^{-2\pi i fT}}{2}\Big[&e^{+i\pi fT_-}\mathrm{sinc}(\pi fT_+)\\
   &+e^{-i\pi fT_+}\mathrm{sinc}(\pi fT_-)\Big],
\end{split}
\label{eq:slowrot:armtransfer}
\end{equation}
with $T_\pm=T\,(1\pm\hat a\!\cdot\!\hat n)$ the light-crossing times for the outbound and inbound
legs, Doppler-shortened and lengthened by the wave's projection onto the arm.  The two
$\mathrm{sinc}$ factors are the averaging on each leg; the two phase factors are the offsets
between them.  A Michelson interferometer differences its arms, so the detector's response to
polarization $A\in\{+,\times\}$ is
\begin{equation}
F_A(f)=\tfrac12\big[\tilde D(\hat x,\hat n,f)\,\varepsilon^A\!:\!\hat x\hat x
   -\tilde D(\hat y,\hat n,f)\,\varepsilon^A\!:\!\hat y\hat y\big],
\label{eq:fsize:FA}
\end{equation}
for arm directions $\hat x,\hat y$ and polarization basis tensors
$\varepsilon^A(\alpha,\delta,\psi)$.  As $f\to0$ every $\mathrm{sinc}\to1$ and
Eq.~\eqref{eq:fsize:FA} collapses to the long-wavelength antenna pattern of
Section~\ref{sec:sub:L}; the first null is at the free spectral range $f_{\rm FSR}=c/2L$
($3747\,\mathrm{Hz}$ for a $40$-km Cosmic Explorer arm).  For today's $\sim\!4$-km detectors the
in-band distortion is negligible; for the $10$--$40$-km arms of the next generation it is a
first-order, in-band effect.
Equation~\eqref{eq:fsize:FA} is a genuine obstruction to the factored likelihood, not a
bookkeeping nuisance.  In Eq.~\eqref{eq:resp:static:ip} the antenna factor was a constant that
passed through the inner products as a prefactor; it is now a frequency-domain multiplication
operator sitting \emph{inside} every noise-weighted integral, and it depends on the sky.  Were we
to leave it there, $Q$, $U$ and $V$ would acquire a sky dependence and would have to be recomputed
at every extrinsic sample --- which is precisely the amortization that makes RIFT viable.  The
remedy is the one used for the rotation in Section~\ref{sec:slowrot}: write the offending operator
as a short sum of sky-analytic scalars times sky-independent operators.  Two observations do it.
\emph{First, the direction-independent part of the transfer is a pure translation.}  Both arms
share the prefactor $e^{-2\pi ifT}$ of Eq.~\eqref{eq:slowrot:armtransfer}, which by
Eq.~\eqref{eq:resp:translation} is $\mathcal{T}_T$: a rigid shift of the template by one
light-crossing time ($1.3\times10^{-4}\,\mathrm{s}$ for a $40$-km arm).  Being common to both
arms it survives the difference in Eq.~\eqref{eq:fsize:FA}, and being a rigid time shift it is
degenerate with the coalescence time, so the time marginalization already present absorbs it
exactly --- no approximation, and no new parameter.  Removing it leaves a residual transfer
\begin{equation}
\hat D(u,x)\equiv e^{+2\pi ifT}\,\tilde D,\qquad u\equiv\hat a\!\cdot\!\hat n,\quad x\equiv\pi fT,
\label{eq:fsize:Dhat}
\end{equation}
which is the whole point of the split: $\hat D$ depends on the sky \emph{only} through the single
scalar $u$ and on frequency \emph{only} through the light-crossing phase $x$, with $\hat D\to1$ as
$x\to0$.  The sky and the frequency have been separated into different arguments of one function
of two variables.
\emph{Second, that residual separates exactly in $u$.}  Because $\mathrm{sinc}$ is an entire
function, $\hat D(u,x)$ is entire in $u$ at fixed $x$, so its Taylor series about $u=0$ converges
for every direction on the sky --- not merely for near-normal incidence:
\begin{equation}
\hat D(u,x)=\sum_{p\ge0}u^{\,p}\,W_p(x),\qquad
W_p(x)\equiv\frac{1}{p!}\,\frac{\partial^{\,p}\hat D}{\partial u^{\,p}}\bigg|_{u=0} .
\label{eq:fsize:Wp}
\end{equation}
Each coefficient $W_p$ is a function of frequency alone, fixed once the detector geometry is
known; each power $u^p$ carries the sky.  Because the two arms of a detector share a common $T$,
they share the same $\{W_p\}$ and differ only through $u$, so a single frequency basis serves the
whole instrument.  Substituting Eq.~\eqref{eq:fsize:Wp} into Eq.~\eqref{eq:fsize:FA} and
collecting the complex combination $F_k=F_+-iF_\times$ then gives
\begin{equation}
F_k(\alpha,\delta,\psi,f)=e^{-2\pi ifT}\sum_{p\ge0}
   b^k_p(\alpha,\delta,\psi)\,W_p(\pi fT),
\label{eq:fsize:factorization}
\end{equation}
\begin{equation}
b^k_p=\tfrac12\big[(\hat x\!\cdot\!\hat n)^p\,\varepsilon\!:\!\hat x\hat x
   -(\hat y\!\cdot\!\hat n)^p\,\varepsilon\!:\!\hat y\hat y\big],
\label{eq:fsize:bp}
\end{equation}
with $\varepsilon=\varepsilon^+-i\varepsilon^\times$ the complex polarization tensor whose
contraction gives the complex response of Section~\ref{sec:sub:L}.  The $p=0$ term is worth
naming: $W_0(x)=\mathrm{sinc}(x)\cos x$ is the on-axis averaging common to all directions, and
$b^k_0$ is exactly the long-wavelength antenna pattern, so the leading term of
Eq.~\eqref{eq:fsize:factorization} \emph{is} the familiar static response with a single
sky-independent frequency taper.  Successive terms add the direction dependence of the averaging,
weighted by increasing powers of the projection $u$.
Two properties of this expansion deserve emphasis, because they differ from the rotation case.
It is a reordering rather than an expansion in a small parameter: convergence is guaranteed for
all $u$, and the order $p_{\max}$ needed for a given accuracy is set by the largest in-band
light-crossing phase $\pi f_{\max}L/c$, so it grows with arm length and with the upper frequency
retained rather than with any property of the source.  And the weights are Hermitian:
$\hat D(u,-x)=\hat D(u,x)^*$, so $W_p(-f)=W_p(f)^*$, the weighted templates remain the transforms
of real timeseries, and the quadratic term needs none of the harmonic reflection that
Eq.~\eqref{eq:slowrot:QUV} requires.
The payoff is that Eq.~\eqref{eq:fsize:factorization} has exactly the structure of
Eq.~\eqref{eq:slowrot:master}, with the frequency weights $W_p$ playing the role of the modulation
operators $\mathcal{M}_n$ and the scalars $b^k_p$ the role of $C^k_a$.  The $W_p$ are folded once
into the frequency-domain modes, $\tilde h_{\ell m}(f)\mapsto W_p(\pi fT)\,\tilde h_{\ell m}(f)$,
and the resulting bank feeds the mode-inner-product primitives of Eq.~\eqref{eq:QUV} unchanged.
The likelihood is again Eqs.~\eqref{eq:slowrot:lnL:Decomposed}--\eqref{eq:slowrot:lnL:MatrixForm}
with the composite index $A=(p,\ell m)$ and $b^k_p$ in place of $\tilde C^k_a$, so $Q$, $U$ and
$V$ remain sky-independent and the extrinsic contraction remains a scalar sum; truncating at
$p_{\max}=0$ recovers the standard likelihood exactly.  We refer to this generalization as
\emph{Path~D}.
The truncation is the one approximation in Path~D, and it is least attractive exactly where the
signal content is broadest --- strongly precessing sources with significant higher multipoles,
whose power extends to the highest in-band frequencies.  For those an alternative route retains
the full frequency-dependent response $F_k(f)$ at a \emph{fixed} sky location, evaluating
Eq.~\eqref{eq:resp:static:ip} directly; this trades the sky's status as an extrinsic parameter
for exactness, and is the appropriate choice once a coarse localization is already in hand.
The physical size of the effect is set by the in-band light-crossing phase, and it is not
subtle at third-generation arm lengths.  On a finite-size injection at fixed loud SNR, the
$\lnL$ recovered by the finite-size likelihood exceeds that of the long-wavelength model by
$0.25$ for a $4$-km LIGO arm --- a null result, as expected --- and grows monotonically with arm
length to $39.6$ for a $40$-km Cosmic Explorer, with $p_{\max}$ raised as $fL/c$ grows.
Both generalizations are implemented in RIFT's vectorized extrinsic integrator, reusing the
existing mode-inner-product and marginalization code paths without modification.
\section{Sampling the differentiable extrinsic likelihood}
\label{sec:jax_ile}
\label{sec:jax_ile:sampling}
Sections~\ref{sec:response} and~\ref{sec:quadrature} define the likelihood evaluated at each
intrinsic point and integrate the extrinsic coordinates that admit direct marginalization.  Let
\(\bm\theta\) denote the remaining sampled coordinates and
\(\mathcal L_{\rm dir}(\bm\lambda,\bm\theta)\) the likelihood after those inner integrals.  The
extrinsic sampler must provide both draws from
\begin{equation}
p(\bm\theta\mid\bm\lambda,d)
=\frac{\mathcal L_{\rm dir}(\bm\lambda,\bm\theta)
p(\bm\theta)}{\mathcal Z(\bm\lambda)}
\end{equation}
and the normalization
\(\mathcal Z(\bm\lambda)=\int d\bm\theta\,
\mathcal L_{\rm dir}p(\bm\theta)\), which is the marginal likelihood consumed by RIFT's intrinsic
stage in Eq.~\eqref{eq:lnL:MonteCarlo}.
This target becomes narrow and multimodal as the network amplitude grows.  Under regular local
curvature, each well-constrained width scales as \(1/\rho\); if a product proposal remains broad
in \(d\) such directions, its overlap with the posterior scales as \(\rho^{-d}\).  Correlations
among sky position, orientation, and arrival time further reduce the usefulness of a separable
proposal.
Two transformations simplify the sampled geometry.  Network-frame sky coordinates rotate the
celestial sphere so that a constant time-delay ring is a constant-polar-angle curve; the rotation
preserves the sky measure.  For quadrupole-dominated signals,
\(\phi_\pm=\phi_{\rm ref}\pm\psi\) aligns the leading phase--polarization degeneracy with coordinate
axes.  Alternatively, the direct treatments of Section~\ref{sec:quadrature} remove
\(\phi_{\rm ref}\) and \(\psi\) before sampling.  These operations reduce the correlations or
dimensions presented to a sampler, but they do not by themselves ensure that every remaining sky
mode is explored.
\subsection{Mode coverage, adaptive clouds, and evidence}
\label{sec:jax_ile:samplers}
We use two complementary strategies.  Multi-start NUTS~\cite{Hoffman2014NUTS} applies the
network-frame rotation and seeds one Hamiltonian chain for each sky mode found by a pilot prior
scan.  The pooled chains define a Gaussian-mixture importance proposal for the normalization.
The second strategy integrates the phase-like degeneracy and evolves a cloud of walkers.  The
complete combinations of sampled coordinates, direct marginalizations, and evidence estimators
are listed in Appendix~\ref{ap:sampler_menu}.
For the cloud strategy we use adaptive sequential Monte Carlo
\cite{Neal2001AIS,DelMoral2006SMC}.  It connects the prior to the posterior through
\begin{equation}
\pi_{\beta_k}(\bm\theta)
=\frac{\mathcal L_{\rm dir}(\bm\lambda,\bm\theta)^{\beta_k}
p(\bm\theta)}{\mathcal Z_{\beta_k}},
\qquad 0=\beta_0<\cdots<\beta_K=1 .
\end{equation}
At rung \(k\), walkers from \(\pi_{\beta_{k-1}}\) receive incremental weights
\begin{equation}
\widetilde w_i^{(k)}
=w_i^{(k-1)}
\mathcal L_{\rm dir}(\bm\lambda,\bm\theta_i)^{\beta_k-\beta_{k-1}},
\qquad
\frac{\widehat{\mathcal Z}_{\beta_k}}
{\widehat{\mathcal Z}_{\beta_{k-1}}}
=\sum_i\widetilde w_i^{(k)} .
\end{equation}
The next \(\beta_k\) is chosen to control the effective sample size of these weights.  After
resampling, random-walk Metropolis moves use a covariance estimated from the current cloud.  The
proposal scale therefore contracts only after walkers have reached the corresponding posterior
region, rather than being fixed by a global fit made before the peak is covered.
The tested normalizing-flow and curvature-preconditioned configurations instead commit to one
fitted global geometry.  In our internal tests where we varied source amplitude, those
configurations lose sample diversity as the target narrows, whereas the adaptive cloud retains
diverse draws.  This is a result about the tested
configurations, not a universal limitation of flows or Hamiltonian methods.  The same sequence does not establish the correctness of an evidence after
sample collapse: no evidence is reported from it, because both arms
used direct-marginalization settings that were later superseded.
For a converged cloud we also form a defensive importance proposal
\(q(\bm\theta)=\alpha p(\bm\theta)+(1-\alpha)
\mathcal N(\bm\theta;\widehat{\bm\mu},s\widehat{\bm C})\), where
\(\widehat{\bm\mu}\) and \(\widehat{\bm C}\) are the cloud moments and \(s>1\) inflates its tails.
An independent draw \(\bm\theta_i\sim q\) then gives
\begin{equation}
\widehat{\mathcal Z}
=\frac{1}{N}\sum_{i=1}^{N}
\frac{\mathcal L_{\rm dir}(\bm\lambda,\bm\theta_i)
p(\bm\theta_i)}{q(\bm\theta_i)} .
\end{equation}
At high amplitude this estimate is well conditioned because its fitted component is built from
samples that already cover the posterior.  At low amplitude a single Gaussian is inadequate for
the multimodal sky, so the cloud-fitted estimate is gated off and the ladder-based value is used
instead.  Multi-start
NUTS uses the same identity with one Gaussian component per recovered mode.
\subsection{Computational cost and proposal reuse}
\label{sec:jax_ile:cost}
The extrinsic stage is evaluated once for every intrinsic template, so its per-point cost is
multiplied across the intrinsic grid.  For the samplers used here we measure
\(\sim\jxWallLo\)--\(\jxWallHi\) minutes per intrinsic point on one modern GPU after the one-time
compilation of the fused kernels.  The high-amplitude cloud is more expensive than the broad-regime
samplers because it carries many walkers through a tempering sequence.  This motivates reuse
across nearby intrinsic points, but the reusable object must preserve posterior coverage.
Let \(\mathcal C_{j-1}\) be an exactly weighted fair draw from intrinsic point \(j-1\).  At the next
point we rebuild a kernel-density proposal from that cloud and combine it with the prior:
\begin{equation}
\begin{aligned}
q_j(\bm\theta)
 &=\alpha\,p(\bm\theta)+(1-\alpha)\,
   \mathrm{KDE}\!\left(\mathcal C_{j-1}\right),\\
w_i
 &=\frac{\mathcal L_{\rm dir}(\bm\lambda_j,\bm\theta_i)
   p(\bm\theta_i)}{q_j(\bm\theta_i)} .
\end{aligned}
\label{eq:cloud_reuse}
\end{equation}
The proposal is refitted at every intrinsic point; only corrected samples are carried forward.
The prior component provides defensive support, and the exact weight removes dependence of the
target on the quality of the density fit.  By contrast, reusing an uncorrected flow can contract
the reported posterior, while correcting a poorly matched reused flow preserves the target but
leaves too few effective samples.
The tested reuse experiment is retained in the internal research notes.  More widely separated
intrinsic points require a fresh coverage check because the prior component can contribute little
probability to a sharply localized sky mode.

\section{Core single-event demonstrations}
\label{sec:results_core_3g}
In this section, we demonstrate our work with  handful of heterogeneous examples.
Section \ref{sec:results:bns} demonstrates a prosaic application -- end-to-end analyses for contemporary ground-based
instruments, without including sophisticated 3G detector response -- using JAX-ILE as a drop-in replacement for existing code.
Conversely, Section \ref{sec:3g_demo} validates (two variants of) JAX-ILE specifically in a high-SNR 3G configuration,
including multiple detectors and full detector response.
Section~\ref{sec:3g_bns_pe} applies the full iterative workflow to a lower-amplitude
3G BNS and presents a joint intrinsic--extrinsic posterior.
\subsection{Full intrinsic/extrinsic analyses with contemporary sources}
\label{sec:results:bns}
\begin{figure}
\includegraphics[width=\columnwidth]{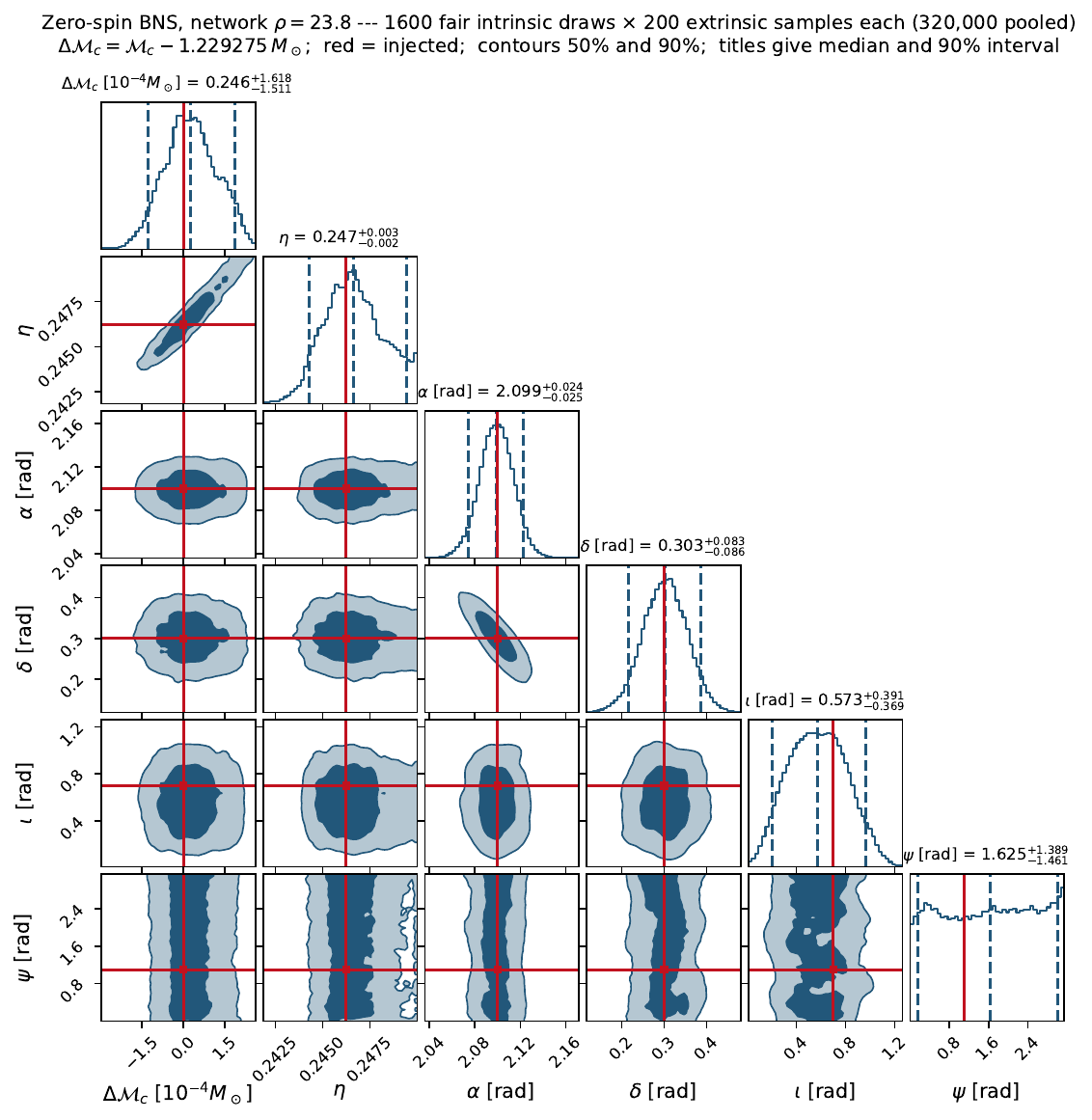}
\caption{\label{fig:bns_joint}\textbf{The differentiable likelihood closes a complete RIFT analysis
for the control binary neutron star.} The injection uses zero noise, an H1--L1--V1 design network,
the static long-wavelength response, and a point-particle dominant-quadrupole waveform. All
$\bnsNparam$ sampled parameters are pooled from $\bnsNdraw$ fair draws of
the intrinsic posterior, each carrying its own cloud of $\sim\!\bnsNext$ extrinsic samples
($\bnsNpooled$ pooled in total); contours enclose $50\%$ and $90\%$, and injected values are red.
Chirp mass is drawn as an offset because its posterior is of order $10^{-5}$ in relative width.
Every parameter is recovered within $\bnsWorstBias\,\sigma$, with the injection inside the $90\%$
credible interval in all $\bnsNInside$ cases, and the sky localized to
$\bnsSkyArea\,\mathrm{deg}^2$ at $90\%$.  This analysis uses the JAX ILE driver with its
phase-marginalized flowMC sampler; luminosity distance is marginalized analytically and is not a
sampled coordinate.}
\end{figure}

\begin{figure*}
\centering
\includegraphics[width=0.72\textwidth]{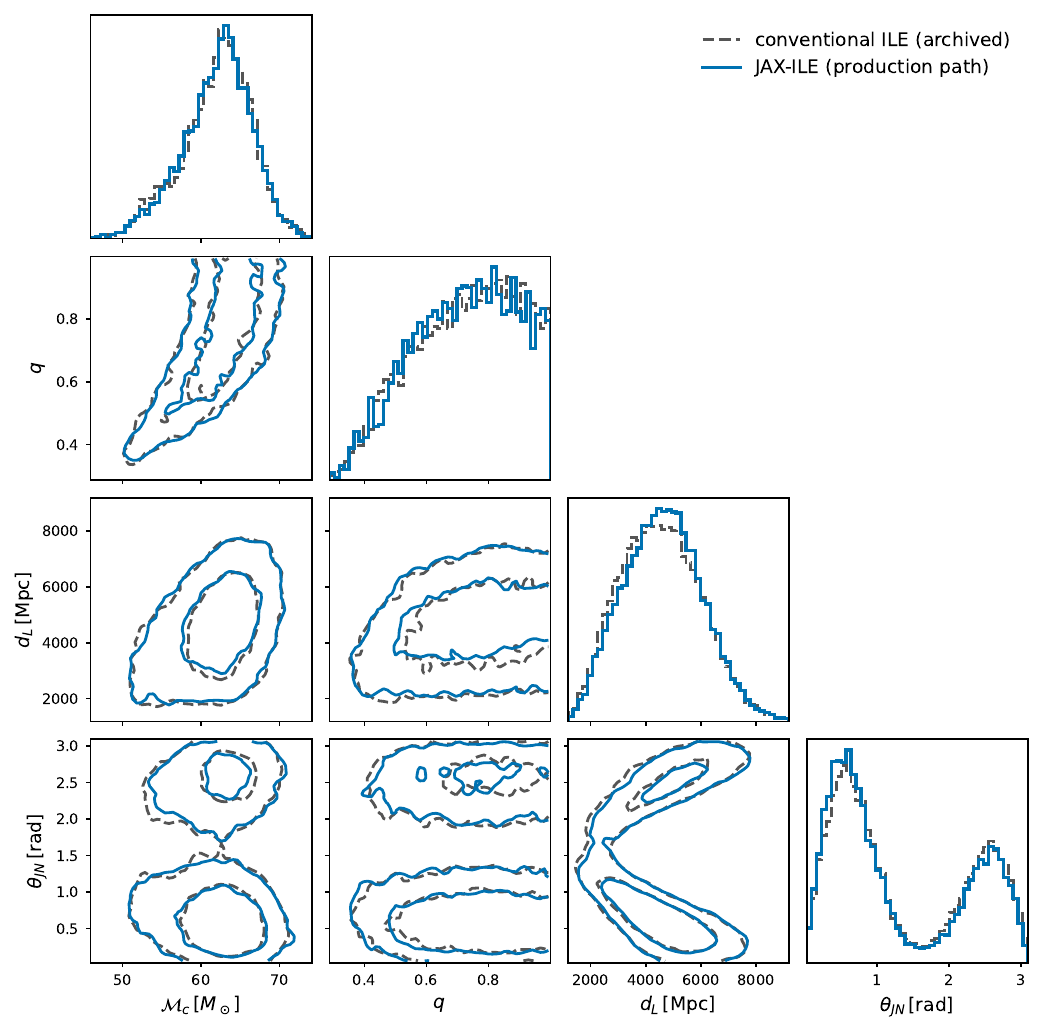}
\caption{\label{fig:bns_gw240426_jax}
\textbf{Two zero-spin analyses of GW240426 give closely overlapping posteriors.}
One- and two-dimensional marginals for detector-frame chirp mass, mass ratio, luminosity distance,
and inclination, using observed H1--L1 strain. Gray dashed curves show conventional ILE; blue
solid curves show JAX-ILE. Both analyses use the same data, noise spectra, waveform model,
priors, and initial intrinsic grid. A same-grid repeat agrees within three combined Monte Carlo errors
for $\jaxEERepeatLnLWithinThreeSigmaPct\%$ of $\jaxEEIntrinsicPoints{}$ likelihood values.}
\end{figure*}

In this section, we demonstrate
whether the
differentiable likelihood can replace the production extrinsic stage within the full iterative
pipeline of Section~\ref{sec:review}, including interpolation, posterior generation, and the
fair-draw extrinsic pass.
For simplicity, we do so using a highly restrictive environment: contemporary ground-based networks, without
incorporating the response function improvements described in Section~\ref{sec:response}, and with zero spin.
We do so  with a controlled binary-neutron-star injection, then
with a matched pair of zero-spin analyses of GW240426 in observed H1--L1 data.  The injection uses
the JAX ILE driver with its phase-marginalized flowMC mode; the GW240426 comparison instead holds
the adaptive-volume (AV) sampler fixed while changing from conventional GPU-vectorized ILE to
JAX-ILE (AV-JAX).
Figure~\ref{fig:bns_joint} shows the results of analyzing a zero-spin, point-particle binary neutron star
($m_1=\bnsMOne\,M_\odot$, $m_2=\bnsMTwo\,M_\odot$; $\mc=\bnsMc\,M_\odot$, $\eta=\bnsEta$) modeled with IMRPhenomD, injected
at $\bnsDist\,\mathrm{Mpc}$ into zero noise in the H1--L1--V1 network, using the Advanced LIGO
design power spectral density for H1 and L1 and the same curve as a simplified V1 design model.
This gives a network amplitude $\rho=\bnsSnr$, and we analyze the signal from
$\bnsFmin\,\mathrm{Hz}$.  The intrinsic parameters are the two mass coordinates.
Its extrinsic stage runs
\texttt{integrate\_likelihood\_extrinsic\_jax} in \texttt{flowmc-phimarg} mode, using flowMC with
phase marginalization rather than AV.  The analysis ran
$\bnsNiter$ iterations of $\bnsNpts$ intrinsic points, each iteration followed by a perturbed
resampling of the current posterior, accumulating $\bnsNile$ likelihood evaluations.  A final
pass expanded $\bnsNdraw$ fair draws from the intrinsic posterior into $\sim\!\bnsNext$ extrinsic
samples apiece.
For a second full analysis, we use the no-calibration, zero-spin GW240426 configuration.  We compare
an archived conventional GPU-vectorized ILE analysis using AV with a JAX-ILE analysis also using
AV (AV-JAX).  The intended science-level change is the likelihood implementation; the AV-JAX arm
also requires its own compilation cache, fixed-shape batching, and fair-draw assembler.  Both
analyses use the same 8~s of H1--L1  strain and
per-detector PSDs, sampled at 1024~Hz with a requested 4096~Hz internal likelihood cadence, a
20--448~Hz analysis band, a nonspinning \textsc{IMRPhenomXPHM} waveform through $\ell=3$, a
Euclidean distance prior, nearest-sample time interpolation, and the same mass and distance bounds.
They also share the archived 500-point initial intrinsic grid, random-forest intrinsic fit, an
eight-iteration request with terminal-stage convergence, adaptive-volume sampler with an
effective-sample target of 100, and 20 intrinsic points per likelihood worker.  Their final tables
represent the same four displayed coordinates with five conditional fair draws at each of
$\jaxEEIntrinsicPoints{}$ intrinsic points.  The 20-point workers and target of 100 are retained to
make the science-level comparison controlled, not as a recommended production operating point.
Figure~\ref{fig:bns_gw240426_jax} shows the resulting intrinsic and extrinsic marginals.  The largest
one-dimensional Jensen--Shannon divergence among the four displayed parameters is
$\jaxEEMaxJS$ bits.  Because the two complete analyses use independently generated stochastic
intrinsic grids, this residual includes both extrinsic-integration and iterative-fit variation.
We therefore also reran JAX-ILE on the same final grid: $\jaxEERepeatLnLWithinThreeSigmaPct\%$ of the
$\jaxEEIntrinsicPoints{}$ paired likelihood values agree within three combined Monte Carlo errors,
and the independently redrawn distance and inclination marginals pass their finite-sample tests.
Together with the injected-signal recovery above, this comparison demonstrates that the
differentiable likelihood operates inside both a controlled end-to-end recovery and a production
workflow on contemporary detector data.
\subsection{Injection--recovery for loud third-generation sources}
\label{sec:3g_demo}
\begin{figure*}
\centering
\includegraphics[width=0.88\textwidth]{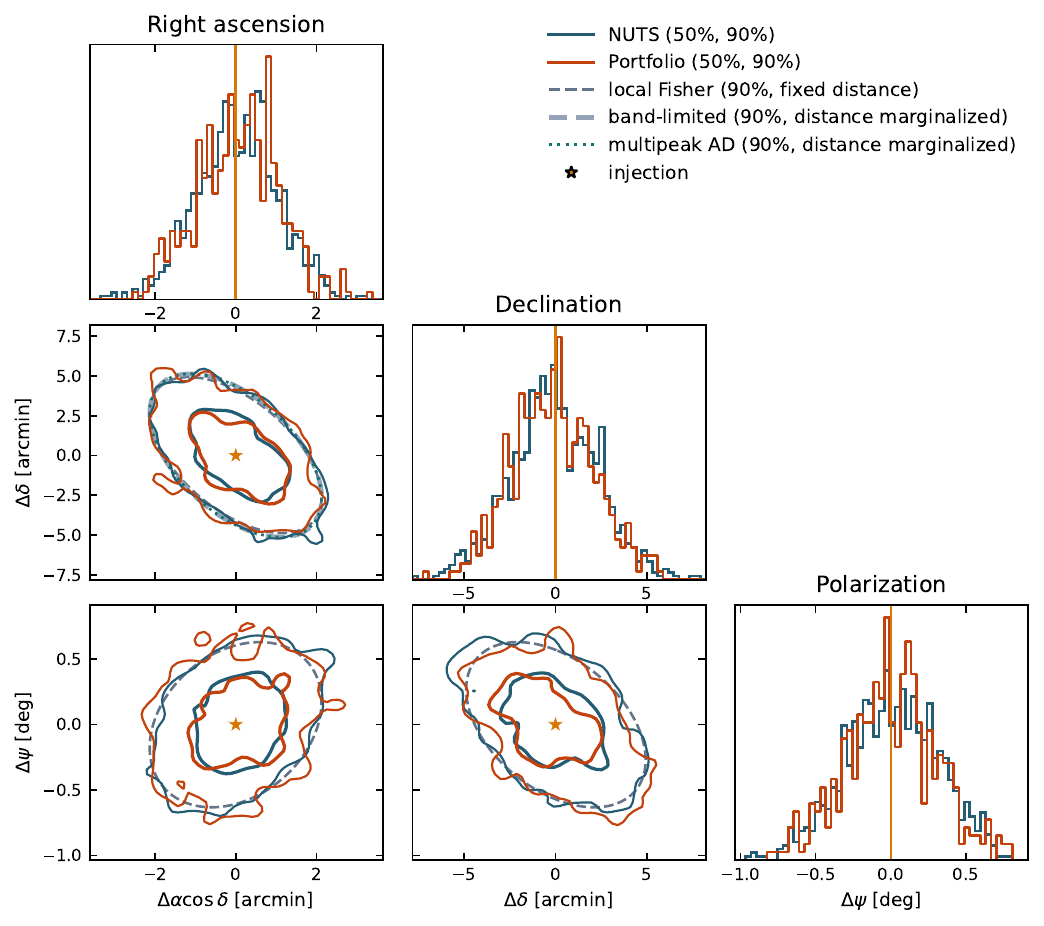}
\caption{\label{fig:3g_recovery}\textbf{Independent Fisher constructions agree, and both
posterior samplers meet the fixed-distance target.}  The marginal posterior is shown for the
finite-size binary-neutron-star injection at network $\mathrm{SNR}=\jsrSnr$ in the CE $+$ ET
$+$ K network.  Blue and red curves enclose \jsrCredibleInnerPct\% and
\jsrCredibleOuterPct\% for NUTS and the value-only AV+GMM portfolio, respectively; the orange
markers denote the injection.  In the sky panel, the light-gray ellipse comes from the guarded
band-limited likelihood, while the dotted teal ellipse comes from the independently evaluated
bounded multipeak likelihood.  Both are marginalized over luminosity distance.  The dashed gray
ellipse shows the band-limited result with distance instead held fixed, as in the posterior
recoveries.  Coordinates are offsets from the injected right ascension, declination, and
polarization; the right-ascension coordinate is $\Delta\alpha\cos\delta$, and polarization is
wrapped modulo $\pi$.  Inclination and reference phase are marginalized from the displayed
posterior.}
\end{figure*}

\begin{figure}
\centering
\includegraphics[width=\columnwidth]{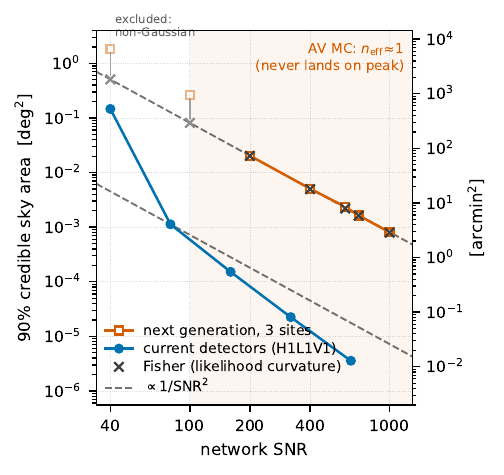}
\caption{\label{fig:3g_skyarea}\textbf{The credible sky area contracts with amplitude and agrees
with the local curvature in the Gaussian regime.}  The blue series shows the current-detector
injection of the amplitude sweep in the Supplemental Material, recovered with adaptive SMC; the
orange series shows the finite-size binary-neutron-star injection in the three-site
next-generation network, recovered with NUTS.  Crosses denote the corresponding local Fisher
areas, dashed lines show $\mathrm{SNR}^{-2}$ scalings, and the shaded band marks the regime in
which the production adaptive Monte Carlo calculation has $n_{\rm eff}\!\approx\!1$.  The right
axis gives the same areas in $\mathrm{arcmin}^2$.}
\end{figure}

Third-generation observatories such as Cosmic Explorer and the Einstein
Telescope~\cite{Evans2021CosmicExplorer,LIGO-Detector-Planning3d-Evans2016} will observe
binary-neutron-star signals with durations of minutes to hours and network signal-to-noise ratios
of several hundred or more.  The duration requires a time- and frequency-dependent detector
response, while the large amplitude concentrates the extrinsic likelihood into a small angular
region~\cite{2025PhRvD.112j2004B}.  We combine the finite-size response of
Section~\ref{sec:response} with the differentiable extrinsic likelihood of
Section~\ref{sec:jax_ile} and test the resulting posteriors in two ways: gradient-based NUTS uses
automatic differentiation (AD), while an adaptive-volume plus Gaussian-mixture (AV+GMM)
portfolio uses the same JAX likelihood only for its values.  We likewise calculate the local
Fisher geometry by differentiating two independently evaluated forms of the likelihood.
We inject a nonspinning binary neutron star into a three-site network comprising a $40$-km Cosmic
Explorer detector, the $10$-km Einstein Telescope triangle, and a third CE-class detector at a
well-separated site.  Each detector uses the finite-size (Path~D) response.  We vary the source
distance to produce network amplitudes from $\tgSnrLo$ to $\tgSnrHi$.  The amplitude sweep uses
an analysis band extending to $\fvFmax\,\mathrm{Hz}$ and a mode-correlation sampling rate of
$\fvSrate\,\mathrm{Hz}$; the detailed posterior comparison uses the
$\mathrm{SNR}=\jsrSnr$ realization with a $\jsrRateHz\,\mathrm{Hz}$ correlation grid.
Before comparing samplers, we establish the local Gaussian target that they must reproduce.  For
consistency with the data handling in the recovery, both Fisher calculations use
finite-difference Hessians of AD likelihood gradients.  The first uses the guarded band-limited
time integral while retaining polarization, inclination, reference phase, and distance as
coordinates; these nuisance directions are marginalized in the inverse Fisher matrix.  The
second directly marginalizes time, reference phase, polarization, and distance with the bounded
multipeak, peak-local policy of Table~\ref{tab:scaling_plan}, and differentiates the resulting
likelihood in sky position and inclination.  The corresponding distance-marginalized sky
ellipses in Figure~\ref{fig:3g_recovery} coincide at plotting precision: the three-dimensional
marginal covariances agree to \madCovDiffPct\% in relative eigenvalues.  This agreement validates
the curvature calculation; because the posterior demonstrations hold distance fixed, their target
is the fixed-distance specialization of the band-limited calculation.
We recover the fixed-distance posterior with two complementary methods.  The AD path
applies dense-mass NUTS to the JAX likelihood at the injected distance and samples
$(\alpha,\delta,\psi,\iota,\phi_{\rm ref})$.  The non-AD path passes values of the same likelihood
to a defensive AV+GMM portfolio; its local Fisher geometry initializes the proposal, while the
GMM member retains full-prior support.  Both use network-frame sky coordinates and
$\phi_\pm=\phi_{\rm ref}\pm\psi$ to reduce the leading quadrupole phase--polarization correlation.
Their sky areas differ by \jsrPortfolioNutsDiffPct\% and remain within
\jsrPortfolioFisherDiffPct\% of the fixed-distance Fisher target.
The chains and portfolio are initialized within the correct sky basin.  This setup tests local
posterior resolution rather than all-sky mode discovery.  In an analysis of observed data, an
independent search localization can provide such a proposal without changing either the prior or
the likelihood estimator.  The Fisher calculation is also local: here it is evaluated at the
injected parameters, whereas an analysis of observed data must first locate the relevant mode.
At this amplitude, the native correlation time series does not resolve the time integral.  We
therefore use cubic interpolation at each detector arrival time and guarded spectral refinement
of the complex correlation primitive.  A half-sample translation of the native grid leaves the
local curvature stable; the Supplemental Material reports the convergence sequence.  The
bounded multipeak construction provides a second check because it reaches the same marginal
curvature through a different treatment of the extrinsic coordinates.  Joint distance recovery
and the nonlinear distance-marginalized spectral time integral are not tested here.
Finally, we test how the recovered sky area scales with amplitude.  Figure~\ref{fig:3g_skyarea}
shows that the sampled area follows the expected inverse-square scaling and agrees with the local
Fisher area to \fvAgreePct\% across the reported range.  At lower amplitudes the posterior is too
non-Gaussian for a covariance ellipse to be an adequate representation, so we do not use the
Fisher comparison there.  The amplitude sweep predates the guarded time integration used in the
detailed high-amplitude recovery and is therefore used only for this area-scaling result.

\subsection{End-to-end parameter estimation for a 3G binary neutron star}
\label{sec:3g_bns_pe}
\begin{figure*}
\centering
\includegraphics[width=0.94\textwidth]{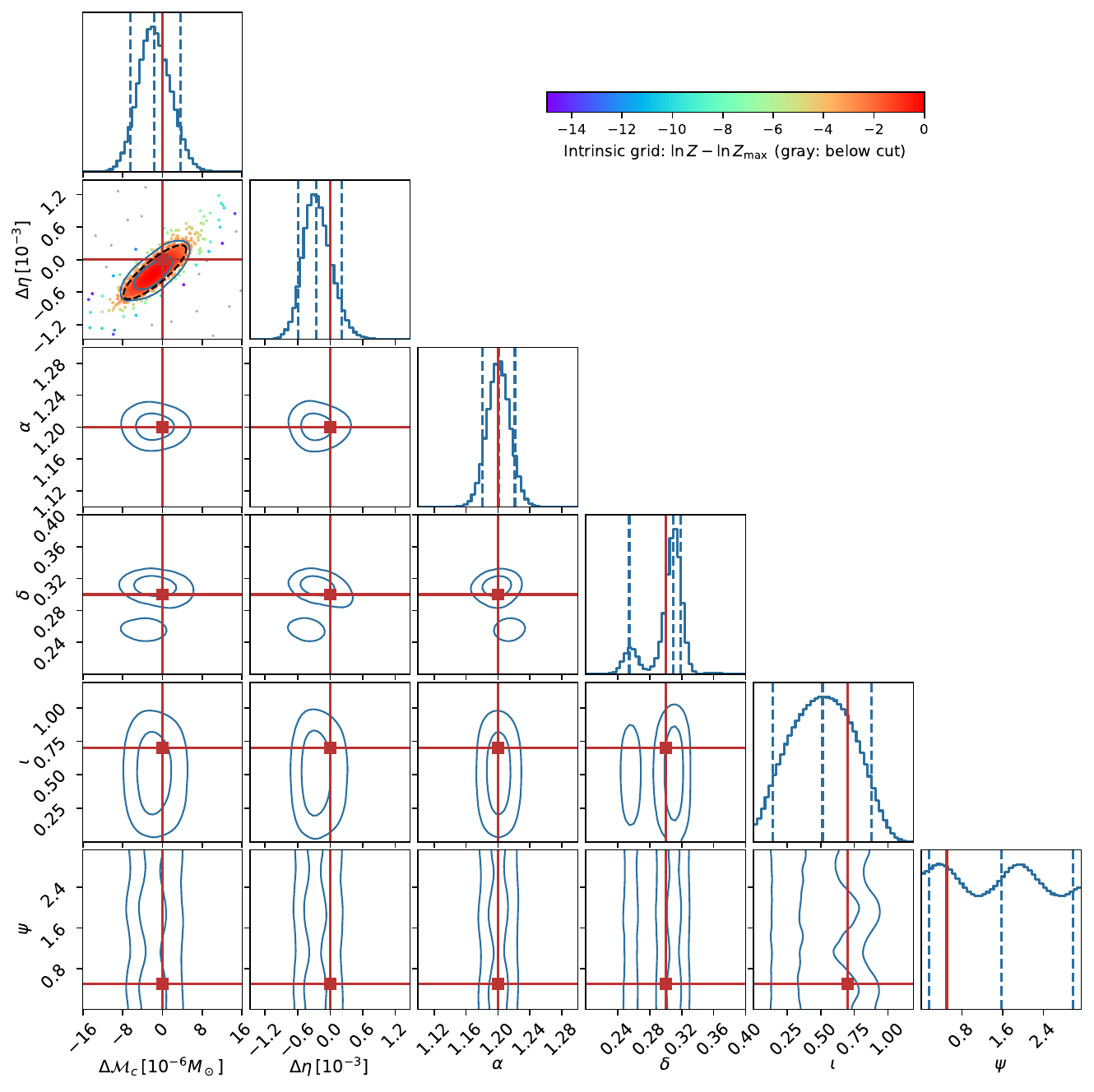}
\caption{\label{fig:3g_bns_joint_pe}\textbf{Joint intrinsic and extrinsic
posterior for the CE--ET--CE BNS at network $\rho=\bnsThreeSnr$.}
The six displayed coordinates are chirp mass and symmetric mass ratio,
right ascension, declination, inclination, and polarization; angles are in
radians. Chirp mass and mass ratio are plotted relative to their injected
values. Blue contours enclose $\bnsThreeInnerContourPct\%$ and
$\bnsThreeCrediblePct\%$ of the weighted joint posterior; red crosshairs mark
the injection. In the intrinsic panel, gray and colored points show evaluated grid locations,
with warmer colors indicating higher marginal likelihood. The posterior uses
$\bnsThreeTerminal$ fair draws and $\bnsThreePool$ pooled conditional draws. The black dashed curve
is the empirical Fisher 90\% ellipse; both intrinsic axes span the full stated prior.}
\end{figure*}

We now combine the finite-size, time-dependent response with the iterative RIFT
parameter-estimation workflow. The controlled source is a nonspinning
$\bnsThreeMOne+\bnsThreeMTwo\,M_\odot$ binary neutron star in zero-noise
CE--ET--CE data at network $\rho=\bnsThreeSnr$. The injection uses the compound
$p=2,q=2$ response; the principal recovery uses $p=1,q=1$, with matched
$p=2,q=1$ and $p=1,q=2$ response controls. We analyze
$\bnsThreeFmin$--$\bnsThreeFmax\,\mathrm{Hz}$ data sampled at
$\bnsThreeRate\,\mathrm{Hz}$ and include the quadrupole modes. Luminosity
distance is numerically marginalized in the extrinsic likelihood, so the
exported joint draws do not include a distance coordinate.
Following the full-pipeline BNS analysis of Section~\ref{sec:results:bns}, an
analytic intrinsic Fisher estimate initializes a grid in chirp mass and
symmetric mass ratio. Successive intrinsic likelihood evaluations, held-out
checked GP fits, posterior-grid draws, and puffball proposals track the
high-likelihood region. The BNS GP controls vary the lower length-scale rule
under the broadened amplitude, noise, and upper length-scale bounds discussed
in the Supplemental Material; the arm with lower held-out error is selected
at each iteration. The final GP proposes $\bnsThreeTerminal$ fair intrinsic
draws. Each receives an independent GPU likelihood integration and conditional
extrinsic sample export. We correct the intrinsic draw weight by the ratio of
measured ILE evidence to the GP value before pooling the joint cloud.
Figure~\ref{fig:3g_bns_joint_pe} provides a corner plot over the non-marginalized intrinsic and extrinsic parameters.
The evaluated marginal likelihood supplies a second, data-driven intrinsic
scale check. We fit a local quadratic in $\mc,\eta$ to ILE values
within $\Delta \lnL\le 3$  of the measured maximum. Its curvature defines an
empirical Fisher matrix over $\mc$, $\eta$ after
extrinsic marginalization.
Comparing their likelihood falloff and
the empirical Fisher 90\% ellipse with the corrected intrinsic posterior
tests both off-peak GP support and the posterior's local curvature.
The fit uses $\bnsThreeFisherFitPoints$ near-peak ILE evaluations.
This empirical fisher matrix is a good characterization of the local intrinsic posterior.
Its RMS
residual is $\bnsThreeFisherFitRms$ in-sample (and
$\bnsThreeFisherCvRms$  in a seeded five-fold held-out check), while
the Gaussian moment
KL divergence between the posterior and this Fisher estimate is $\bnsThreeFisherMomentKl$ nat.
An independently seeded GP refit to the pooled terminal evidences passes all
predeclared PE-scale stability checks. Its held-out RMS is
$\bnsThreeRefitHoldoutRms$ nat; intrinsic quantiles shift by at most
$\bnsThreeRefitMaxQuantileShiftPct\%$ of the first 90\% width and widths by at
most $\bnsThreeRefitMaxWidthShiftPct\%$. The smoothed symmetric two-dimensional
KL divergence is $\bnsThreeRefitKl$ nat, smaller than the seeded half-sample
fluctuations. The weaker low-declination lobe carries
$\bnsThreeLowSkyLobePct\%$ of the corrected posterior, compared with
$\bnsThreeLowSkyHalfOnePct\%$ and $\bnsThreeLowSkyHalfTwoPct\%$ in two seeded
intrinsic halves. The evaluated grids enclose $\bnsThreeFirstHullPct\%$ of
the first and $\bnsThreeRefitHullPct\%$ of the refit draws; weighted mass in
the outer tenth of the chirp-mass and mass-ratio priors is
$\bnsThreePriorMcOuterPct\%$ and $\bnsThreePriorEtaOuterPct\%$. Thus the
directly ILE-corrected cloud in Fig.~\ref{fig:3g_bns_joint_pe} is stable at
the stated PE precision without another terminal grid.
For this straw-man calculation, the overall RIFT cost was modest. The median worker cost involves a
$\bnsThreePrecomputeS\,\mathrm{s}$ precompute, a
$\bnsThreeFirstJaxS\,\mathrm{s}$ first JAX batch including compilation,
$\bnsThreeWarmUs\,\mu\mathrm{s}$ per warm likelihood, and
$\bnsThreeWorkerS\,\mathrm{s}$ total worker wall time. These timings exclude
Condor scheduling and input transfer.

\section{Conclusions}
\label{sec:conclude}
We extended RIFT along two axes needed for third-generation analyses.  First, we generalized the
factored likelihood to the slowly time-varying response from Earth rotation and the
frequency-dependent response of finite-size arms.  Both generalizations preserve the sky as an
extrinsic parameter and add only a short inner sum to the precompute-and-marginalize architecture.
Second, we implemented a differentiable backend for RIFT's extrinsic likelihood and applied NUTS,
normalizing-flow, and adaptive sequential-Monte-Carlo samplers through a common interface.   In an conventional mode
without extra direct-quadrature extrinsic marginalization, this likelihood implementation is substantially faster than the previous (cupy) implementation, so should be the preferred RIFT
operating-point choice even with existing non-differentiable samplers.
Differentiation also provides a local Fisher matrix cheaply, supporting existing non-differentiable samplers with
well-targeted initial proposals.
l
Third, we validated our new framework using three key end-to-end tests:
a sequence of short BBH injections  in a contemporary-detector network over a range
of SNRs  to validate the extrinsic samplers' effectiveness;
the controlled binary-neutron-star recovery and matched conventional-AV/AV-JAX zero-spin analyses of
GW240426, which test the full intrinsic/extrinsic workflow without the response extensions;
and recovery of a realistic 3g source using two independent samplers.
This architecture complements methods that reduce likelihood-evaluation cost through reduced-order
quadrature~\cite{2026arXiv260614197G,gw-astro-ReducedOrderQuadraturePE-TiglioEtAl2014,gwastro-pe-ROM-IMRPv2-2016}
or amortize inference with learned posterior models~\cite{2021PhRvL.127x1103D,2025Natur.639...49D}.
RIFT instead evaluates the waveform model during a one-time precompute of the mode overlaps
$\rho_{\ell m}(t)=\langle h_{\ell m}(t)\,|\,d\rangle$ and $\{U,V\}$ of
Section~\ref{sec:sub:L}.  Subsequent extrinsic evaluations contract small arrays of precomputed
quantities, with analytic distance marginalization.  This design does not remove waveform
generation or precompute cost, but it keeps the sampling-stage working set independent of signal
duration and avoids constructing a reduced basis or pretrained posterior model.
A companion paper provides more information about
production operating points and validation tests for interpreting contemporary sources seen in ground-based instruments \cite{gwastro-RIFT-roboto-core}.
The accelerated implementations and accuracy improvements for RIFT demonstrated in this work have wider and immediate
application to the interpretation of the current gravitational wave census.
In this work, we use simplified proof-of-concept demonstrations (nonprecessing sources, with pre-identified coarse
localization) to demonstrate that RIFT's current architecture can be efficiently applied to 3G sources, with the
acceleration improvements identified in this work.  Multiple other studies have demonstrated effective 3G parameter estimation,
including high SNR and realistic detector response, using more realistic precessing sources \cite{2025PhRvD.112j3015S,2025ApJ...987L..17H,2026arXiv260614197G}.
Our framework is nonetheless useful precisely because RIFT can provide an independent validation code, which already
internally self-consistently validates itself against its own autodifferentiated Fisher matrix calculations; employ more
sophisticated waveforms (including additional physics like higher-order-modes, eccentricity, et cetera)  efficiently, enabling robust
investigations of systematics; and integrate within well-established, production-quality data analysis frameworks
without additional user effort.
\begin{acknowledgements}
  ROS acknowledges support from  NSF PHY 2012057, 2309172 and 2206321. ROS also acknowledges assistance from
  Claude/Codex during the implementation, operation, and drafting of the analyses in this manuscript. The authors are
  grateful for computational
  resources provided by the LIGO Laboratory and supported by National Science
  Foundation Grants PHY-0757058 and PHY-0823459. This research has made use of data or
  software obtained from the Gravitational Wave Open Science Center (gwosc.org),
  a service of the LIGO Scientific Collaboration, the Virgo Collaboration, and
  KAGRA. This material is based upon work supported by NSF's LIGO Laboratory
  which is a major facility fully funded by the National Science Foundation, as
  well as the Science and Technology Facilities Council (STFC) of the United
  Kingdom, the Max-Planck-Society (MPS), and the State of Niedersachsen/Germany
  for support of the construction of Advanced LIGO and construction and
  operation of the GEO600 detector. Additional support for Advanced LIGO was
  provided by the Australian Research Council. Virgo is funded, through the
  European Gravitational Observatory (EGO), by the French Centre National de
  Recherche Scientifique (CNRS), the Italian Istituto Nazionale di Fisica
  Nucleare (INFN) and the Dutch Nikhef, with contributions by institutions from
  Belgium, Germany, Greece, Hungary, Ireland, Japan, Monaco, Poland, Portugal,
  Spain. KAGRA is supported by Ministry of Education, Culture, Sports, Science
  and Technology (MEXT), Japan Society for the Promotion of Science (JSPS) in
  Japan; National Research Foundation (NRF) and Ministry of Science and ICT
  (MSIT) in Korea; Academia Sinica (AS) and National Science and Technology
  Council (NSTC) in Taiwan.
  \end{acknowledgements}
\appendix
\section{Bookkeeping of the sidereal modulation}
\label{ap:slowrot:modulation}
The arrival-time post-phase in Eq.~\eqref{eq:slowrot:postphase} is required because sidereal
modulation and time translation do not commute.  Two bookkeeping errors are especially damaging.
Referencing the modulated templates to the event clock rather than their intrinsic epoch adds a
spurious phase of order $10^4\,\mathrm{rad}$ to the $U,V$ cross terms.  Applying the post-phase to
only one term in $\lnL=\qmstateproduct{d}{h}-\tfrac12\qmstateproduct{h}{h}$ instead evaluates the
two terms with different templates and can violate
$\lnL\leq\tfrac12\qmstateproduct{d}{d}$.
Moving the modulation from the template to the data does not remove this requirement.  If
${\cal S}_n$ shifts frequency by $n\Omega_\oplus/(2\pi)$ and $W$ is the band-limited inverse-noise
weight, the attempted move changes the overlap by
$\langle h|[W,{\cal S}_{-n}]d\rangle$.  The shift may be sub-bin, but the residual is controlled by
the variation of $W$, particularly at the band edges; moreover, the $U,V$ terms have no analogous
data-side construction.  The data and quadratic terms must therefore use the same explicitly
modulated template and the same reference-time convention.
Checks at $\Omega_\oplus=0$, at $t_k=t_{\rm ref}$, or against a reference that shares the same
convention are blind to these errors.  We require both the Cauchy--Schwarz bound and agreement with
an independently constructed time-domain model.  In the stress test described in the repository's
provenance record, the inconsistent implementation exceeded the bound by $0.07548$ nats; after the
fix it remains below the bound by $0.00078$ nats, a margin independent of the imposed rotation rate
and consistent with the time-grid resolution.
\section{Frequency-domain--native precompute for the modulated templates}
\label{ap:fd_precompute}
The bank $\{\chi^{(a)}_{\ell m}\}$ of Eq.~\eqref{eq:slowrot:master} is realized without ever
materializing a family of time-domain templates.  Each frequency-domain mode
$\tilde h_{\ell m}(f)$ is built once, and both operators of Eq.~\eqref{eq:slowrot:master} are
cheap, exact, per-bin operations on it.  The $p$-th time derivative is the weight
$(s\,2\pi if)^p$, with $s$ fixed by RIFT's FFT convention; the modulation $\mathcal{M}_n$ of
Eq.~\eqref{eq:slowrot:modop} is a frequency shift by $n f_{\rm sid}$, with
$f_{\rm sid}=\Omega_\oplus/2\pi\approx1.2\times10^{-5}\,\mathrm{Hz}$ far below the frequency
resolution $\Delta f$, and is applied exactly as a time-domain linear phase through one FFT round
trip rather than by interpolation on the frequency grid.  The frequency weights $W_p$ of
Eq.~\eqref{eq:fsize:Wp} are applied in the same place and the same way.  The modulated templates
so constructed enter the data term and the cross terms alike, so that every overlap in the bank
refers to the same object; the modulation is deliberately \emph{not} pushed onto the data
instead, because a frequency shift does not commute with the noise weight
(Appendix~\ref{ap:slowrot:modulation}).  All of this feeds the existing mode-inner-product
primitives unchanged.
\section{Choosing response truncation orders from the signal}
\label{ap:response-order}
The analysis-band edge gives a safe but generally wasteful way to choose
$(p_{\max},Q_{\max})$.  A signal-weighted rule follows from the waveform omitted by a candidate
pair.  Let $h$ denote a sufficiently high-order reference response, $h_{PQ}$ its truncation, and
$r_{PQ}=h-h_{PQ}$.  For a zero-noise datum $d=h$, the loss at fixed source parameters is exactly
\begin{equation}
 \ln {\cal L}(h)-\ln {\cal L}(h_{PQ})
 =\tfrac12\qmstateproduct{r_{PQ}}{r_{PQ}}.
 \label{eq:response:truncation-loss}
\end{equation}
The same norm bounds the systematic displacement in the Fisher metric: the squared displacement
is the norm of the projection of $r_{PQ}$ onto the waveform tangent space and therefore cannot
exceed $\qmstateproduct{r_{PQ}}{r_{PQ}}$.  If $\kappa$ is the permitted worst-case displacement in
posterior-standard-deviation units, the least expensive acceptable pair satisfies
\begin{equation}
 \max_{\bm\theta\in\Theta}
 \frac{\qmstateproduct{r_{PQ}}{r_{PQ}}}{\qmstateproduct{h}{h}}
 \leq \frac{\kappa^2}{\rho^2},
 \label{eq:response:order-rule}
\end{equation}
over the extrinsic support $\Theta$.  The required fractional response accuracy consequently
scales as $1/\rho$, while the order depends on the signal, detector PSDs, arm lengths, mode
content, and explored extrinsic region rather than on SNR alone.
Equation~\eqref{eq:response:order-rule} requires no data-dependent likelihood scan.  Embed the
candidate and reference responses in one response-weighted mode bank.  If $\Delta\mathcal C$ is
their difference in analytic coefficients, the residual power is the existing contraction
\begin{equation}
 \qmstateproduct{r_{PQ}}{r_{PQ}}
 =\tfrac12\left[\Delta\mathcal C^\dagger U\Delta\mathcal C
 +\operatorname{Re}\!\left(\Delta\mathcal C^T V\Delta\mathcal C\right)\right].
 \label{eq:response:uv-order-rule}
\end{equation}
The derivative-weighted $U,V$ blocks are noise-weighted frequency moments for slow rotation, and
the $W_q^*W_{q'}$ blocks are their finite-arm counterparts.  Their full contraction preserves
interference among modes and response orders.  A finite scan increases the reference orders until
the last computed shells are negligible relative to the declared budget, then selects the
least-cost passing pair.  The detailed certificate, mode-set rule, paper-example scan, and response-cost
plateaus are recorded in the companion research notes; the current implementation uses a finite
deterministic angular design and therefore reports an estimate, with an explicit unresolved-reference
warning, rather than an analytic supremum over $\Theta$.
\section{Likelihood interpolation at third-generation dynamic range}
\label{ap:gp_scale}
The interpolation stage of Section~\ref{sec:sub:CIP} uses random-forest regression by default for
large production analyses.  Its Gaussian-process option provides a useful smooth model, but its
legacy hyperparameter bounds do not accommodate third-generation likelihood ranges.  The kernel is
\begin{equation}
k = \sigma_n^2 + \sigma_f^2 k_{\rm RBF}(\bm\lambda,\bm\lambda'),
\label{eq:ap:gp:kernel}
\end{equation}
with optimized white-noise amplitude $\sigma_n$, signal amplitude $\sigma_f$, and coordinate
length scales.  The default $\sigma_f^2\leq10$ ceiling was calibrated to contemporary-detector
likelihoods.  At third-generation amplitudes it forces both amplitude parameters to their bounds,
turning a sharp likelihood peak into a low-amplitude surface with a large nugget.
For the zero-spin binary-neutron-star recovery of Section~\ref{sec:results_core_3g}, we fit the same
$\bnsGpNfit$ likelihood evaluations with the default bounds and with the signal-amplitude bound
relaxed by five orders of magnitude.  Five-fold cross-validation gives
\begin{center}
\begin{tabular}{lccc}
\hline\hline
 & default & relaxed & relaxed, \\
 & bounds  & bounds  & $\ell_{\rm max}\!\times\!20$ \\
\hline
fitted $\sigma_f$ (nats)      & $\bnsGpSigmaFDefault$ (at bound) & $\bnsGpSigmaFRelaxed$ & $\bnsGpSigmaFRelaxed$ \\
fitted $\sigma_n$ (nats)      & $\bnsGpSigmaNDefault$ (at bound) & $\bnsGpSigmaNRelaxed$ & $\bnsGpSigmaNRelaxed$ \\
hyperparameters at a bound    & $\bnsGpNsat$                     & $\bnsGpNsatRelaxed$   & $\bnsGpNsatRelaxed$ \\
held-out RMS (nats)           & $\bnsGpHoldoutDefault$           & $\bnsGpHoldoutRelaxed$ & $\bnsGpHoldoutRelaxedLs$ \\
surface peak ($\sigma_{\mc}$) & $\bnsGpPeakDefault$              & $\bnsGpPeakRelaxed$    & $\bnsGpPeakRelaxed$ \\
\hline\hline
\end{tabular}
\end{center}
The relaxed fit improves the held-out error by a factor $\bnsGpAccuracyGain$, leaves no
hyperparameter at a bound, and locates the peak within one fifth of the grid's chirp-mass scale.
Expanding the length-scale ceiling by a further factor of twenty changes neither the fit nor its
held-out score, isolating the fixed amplitude bound as the failure.
The posterior medians are stable across $\bnsGpNrep$ interpolation replicates, but the relaxed fit
widens the $90\%$ credible intervals by $\bnsGpWidthGainMc\%$ in chirp mass and
$\bnsGpWidthGainEta\%$ in symmetric mass ratio.  Without an independent intrinsic-posterior
reference we do not identify either width as exact; the supported conclusion is that a saturated
fit is measurably worse and changes reported uncertainty.  The amplitude bound should scale with
the observed training-data range, and the fitted hyperparameters should be reported with explicit
boundary checks.  Coverage near the maximum remains a separate requirement: additional off-peak
evaluations are needed whenever the fitted maximum lies beyond the sampled grid.
\section{Direct marginalization of the extrinsic parameters}
\label{sec:quadrature}
The loudest 3G sources will have extremelly well-constrained extrinsic parameters, notably including sky location.  In practice, the
brute force Monte Carlo methods over the entire prior volume can have difficulty identifying these extremely narrow
peaks efficiently, most notably including sky location.   If a source is already well-localized, the limitations of
brute-force Monte Carlo can be substantially ameliorated by adopting narrower priors over extrinsic parameters,
principally sky location.  That said, code performance can also be significantly improved by taking advantage of the
increasingly concentrated likelihood directly, performing highly efficient quadrature.  For this reason, in this section
we introduce multiple methods to directly marginalize over several extrinsic parameters (not including sky location).  Because the most efficient
local methods necessariliy break down when the likelihood is not sharply peaked, however, when considering direct
quadrature schemes we must consider multiple methods, balancing accuracy and robustness.    Our discussion therefore
covers multiple methods, with overlapping regimes of utility, and the necessariliy hierarchical and SNR-sensitive policy
required to efficiently and robustly choose between them in practice.
For fixed intrinsic parameters $\bm\lambda$, let $\bm\xi$ denote the extrinsic coordinates sampled
by the outer integrator and let
$\bm\eta=(\phi_{\rm ref},\psi,t,\ln D)$ denote those integrated directly inside each likelihood
evaluation.  The quantity supplied to the sampler is
\begin{equation}
 \begin{aligned}
 {\cal L}_{\rm dir}(\bm\lambda,\bm\xi)
 &=\int d\bm\eta\,p(\bm\eta)\exp[\ell(\bm\lambda,\bm\xi,\bm\eta)]\\
 &\simeq \sum_{\bm k}W_{\bm k}
 \exp[\ell(\bm\lambda,\bm\xi,\bm\eta_{\bm k})],
 \end{aligned}
 \label{eq:direct_marginalization}
\end{equation}
where $\ell=\ln {\cal L}_{\rm full}$, $p(\bm\eta)$ includes the appropriate Jacobians, and
$W_{\bm k}$ are the multidimensional quadrature weights.  Errors in this inner sum propagate into
every subsequent evidence and posterior calculation; drawing more outer samples cannot remove
them.
The high-amplitude scaling follows from the local form of the likelihood.  In dimensionless local
coordinates about a mode $\bm\eta_\star$,
\begin{equation}
 \begin{aligned}
 \ell(\bm\eta)
 &=\ell_\star-\frac{\rho^2}{2}
 g_{ab}\Delta\eta^a\Delta\eta^b
 +O\!\left(\rho^2|\Delta\bm\eta|^3\right),\\
 \sigma_a&\simeq\rho^{-1}\sqrt{(g^{-1})_{aa}},
 \end{aligned}
 \label{eq:direct_marginalization_width}
\end{equation}
where $g_{ab}$ is the amplitude-normalized curvature and $\rho$ is the network signal-to-noise
ratio.  A fixed grid of spacing $h_a$ therefore remains resolved only while $h_a\lesssim\sigma_a$;
over a fixed range $R_a$, its node count must grow at least as
\begin{equation}
 N_a\gtrsim R_a/\sigma_a\propto\rho.
 \label{eq:direct_marginalization_nodes}
\end{equation}
Once this condition fails, Eq.~\eqref{eq:direct_marginalization} can remain smooth while its value is
biased.  Appendix~\ref{ap:fisher_coeff} supplies the coefficients $\sigma_a\rho$ for representative
sources, noise curves, and networks, at the three levels of conditioning the pipeline encounters.
Figure~\ref{fig:quadrature_map} organizes the available remedies by the information used to place
the nodes.  A fixed-grid method increases $N_a$ according to
Eq.~\eqref{eq:direct_marginalization_nodes}.  For the angles, exact-reconstruction adaptive
integration (ERAI) first reconstructs the finite harmonic model exactly
[Eqs.~\eqref{eq:ap:angleharm}--\eqref{eq:ap:nyquist}], then integrates its exponential on an
amplitude-sized arithmetic grid.  The analogous structure-based time method reconstructs the
band-limited correlation timeseries [Eq.~\eqref{eq:ap:sinc}] before integration.  A peak-local
method obtains the mode and curvature from the integrand and integrates in units of its local width,
as in the time windows of Eq.~\eqref{eq:ap:tlocal} and the distance rule of
Eq.~\eqref{eq:ap:dgh}.  The latter two strategies keep the number of expensive likelihood
evaluations fixed or move the amplitude dependence into inexpensive reconstruction arithmetic.
The appropriate construction differs by coordinate.  The angular exponent is an exactly finite
trigonometric polynomial, so its expensive evaluation count is fixed by waveform mode content; a
Laplace treatment of a single dominant polarization mode has an error that decreases as
$O(\rho^{-2})$.  In time, $\ell(t)$ is band limited but $\exp[\ell(t)]$ is not, so resolving the
correlation series does not by itself resolve its integral; the required fixed-grid sample rate is
given by Eq.~\eqref{eq:ap:srate}.  In inverse distance $x=D_{\rm ref}/D$, both the peak and its width
are available analytically [Eq.~\eqref{eq:ap:dpeak}], which makes adaptive Gauss--Hermite
quadrature preferable to a grid spanning the prior.  Appendix~\ref{ap:quadrature} gives the
derivations and the full scheme comparison.
\begin{figure*}[!t]
\centering
\begin{tikzpicture}[
  font=\footnotesize,
  fam/.style={font=\small\bfseries},
  ax/.style={font=\small\bfseries},
  cell/.style={draw, rounded corners=2pt, align=center, inner sep=3pt,
               minimum height=7mm, text width=38mm},
  ship/.style={cell, fill=black!7},
  optin/.style={cell},
  blocked/.style={cell, dashed, text=black!55},
  gap/.style={cell, draw=none, align=center, text=black!70, font=\footnotesize\itshape},
]
\node[fam, text width=40mm, align=center] at (0,1.15) {I.\ Refine the grid};
\node[fam, text width=40mm, align=center] at (4.6,1.15) {II.\ Exploit known structure};
\node[fam, text width=40mm, align=center] at (9.2,1.15) {III.\ Locate the peak first};
\node[font=\scriptsize\itshape, text width=40mm, align=center] at (0,0.62) {information from the data};
\node[font=\scriptsize\itshape, text width=40mm, align=center] at (4.6,0.62) {from analytic structure};
\node[font=\scriptsize\itshape, text width=40mm, align=center] at (9.2,0.62) {from the integrand itself};
\draw[semithick] (-2.9,0.32) -- (11.3,0.32);
\draw[->, black!55, semithick] (-1.35,0.02) -- (11.3,0.02);
\node[font=\scriptsize\itshape, text=black!55, anchor=west, fill=white, inner sep=2pt]
     at (0.5,0.02) {node count set by what does not move with the signal $\longrightarrow$ set by the signal itself};
\draw[semithick] (-2.9,-0.28) -- (11.3,-0.28);
\node[ax, anchor=east, text width=25mm, align=right] at (-2.95,-1.2)
     {$\phi_{\rm ref},\psi$\\[-1pt]\scriptsize angles};
\node[optin]   (a1) at (0,-1.0) {product grid\\[-2pt]\scriptsize legacy replay};
\node[ship]    (a2) at (4.6,-0.8) {ERAI (\texttt{exact}; \textbf{default})\\[-2pt]\scriptsize exact model; adaptive integral};
\node[optin]   (a3) at (4.6,-1.6) {analytic Laplace in $\psi$\\[-2pt]\scriptsize \texttt{auto} candidate; conditional GH};
\node[optin]   (a4) at (9.2,-0.8) {peak-local in $\psi$\\[-2pt]\scriptsize Eq.~\eqref{eq:ap:psicells}; $\phi$ dense};
\node[optin]   (a5) at (9.2,-1.6) {peak-local in both angles\\[-2pt]\scriptsize \texttt{phi-local}; driver selectable};
\draw[black!20] (-2.9,-1.95) -- (11.3,-1.95);
\node[ax, anchor=east, text width=25mm, align=right] at (-2.95,-3.0)
     {$t$\\[-1pt]\scriptsize arrival time};
\node[ship]  (t1) at (0,-2.6) {fixed-spacing Simpson\\[-2pt]\scriptsize Eq.~\eqref{eq:ap:tfixed}; \textbf{default}};
\node[ship]  (t1b) at (4.6,-3.5) {sub-sample interpolation\\[-2pt]\scriptsize nearest $\mid$ cubic $\mid$ \textbf{sinc default}};
\node[optin] (t2) at (4.6,-2.6) {terminal band-limited\\[-2pt]\scriptsize mirrored; Eq.~\eqref{eq:ap:tband}};
\node[optin] (t3) at (9.2,-2.6) {peak-local\\[-2pt]\scriptsize Eq.~\eqref{eq:ap:tlocal}};
\node[gap, draw=black!45, dashed, text width=38mm, inner sep=4pt] (tgap) at (9.2,-3.7)
     {differentiable arm: endpoints\\ collapse to $\ln\mathcal{L}(t)$ before\\ the reconstruction applies};
\draw[black!20] (-2.9,-4.3) -- (11.3,-4.3);
\node[ax, anchor=east, text width=25mm, align=right] at (-2.95,-5.35)
     {$d$\\[-1pt]\scriptsize luminosity distance};
\node[ship]  (d1) at (0,-4.95) {linearly spaced grid\\[-2pt]\scriptsize Eq.~\eqref{eq:ap:dgrid}; \textbf{default}};
\node[optin] (d2) at (4.6,-5.85) {log-uniform, tolerance-sized\\[-2pt]\scriptsize Eq.~\eqref{eq:ap:dlog}};
\node[optin] (d3) at (9.2,-4.95) {adaptive Gauss--Hermite\\[-2pt]\scriptsize Eq.~\eqref{eq:ap:dgh}};
\node[optin] (d4) at (9.2,-5.85) {restricted range\\[-2pt]\scriptsize Eq.~\eqref{eq:ap:dbox}};
\draw[black!20] (-2.9,-6.45) -- (11.3,-6.45);
\node[ax, anchor=east, text width=25mm, align=right] at (-2.95,-7.3)
     {sky\\[-1pt]\scriptsize $\alpha,\delta$};
\node[optin] (s1) at (9.2,-7.3) {restricted range\\[-2pt]\scriptsize Eq.~\eqref{eq:ap:dbox} on the sky};
\node[gap, text width=40mm] at (0,-7.3) {no quadrature: the sky is sampled};
\draw[semithick] (-2.9,-8.15) -- (11.3,-8.15);
\node[anchor=west, font=\scriptsize] at (-2.9,-8.6)
  {\tikz\node[ship, text width=3mm, minimum height=3mm, inner sep=1.5pt]{};\ the default \quad
   \tikz\node[optin, text width=3mm, minimum height=3mm, inner sep=1.5pt]{};\ available \quad
   \tikz\node[blocked, text width=3mm, minimum height=3mm, inner sep=1.5pt]{};\ not available, with the reason that blocks it};
\end{tikzpicture}
\caption{\label{fig:quadrature_map}\textbf{Strategies for direct marginalization of each
extrinsic coordinate.}  The columns group methods by whether they refine a fixed grid, use analytic structure,
or locate a likelihood peak. Filled boxes mark defaults, open boxes available alternatives,
and dashed boxes unavailable combinations. The sky is sampled rather than directly
marginalized. The current differentiable likelihood has no structure-based or peak-local
arrival-time rule.}
\end{figure*}
For validation, each directly integrated coordinate must be refined while all other quadratures and
sample locations are held fixed.  If $\Delta_a$ is the resulting change in log marginal likelihood,
the convergence requirement is axis specific,
\begin{equation}
 \Delta_a\equiv
 \ln {\cal L}_{\rm dir}^{(a,{\rm ref})}-
 \ln {\cal L}_{\rm dir}^{(a,{\rm test})},
 \qquad |\Delta_a|<\epsilon_a,
 \label{eq:direct_marginalization_validation}
\end{equation}
with a tolerance $\epsilon_a$ chosen below the analysis error budget.  Agreement of the final
scalar integral is not a sufficient convergence test: quadrature errors are signed and can cancel
between coordinates.  A reliable convergence assessment must also bound both omitted probability
outside the integrated region and discretization error inside it.
Table~\ref{tab:quadrature_menu} in Appendix~\ref{ap:quadrature} summarizes the implementation
choices and their scaling.
\subsection{Error and cost scaling}
\label{sec:quadrature:scaling}
Equations~\eqref{eq:direct_marginalization_width} and
\eqref{eq:direct_marginalization_nodes} give a common origin for the asymptotic cost laws.  If
${\cal D}_q$ is the set of coordinates that a scheme $q$ still resolves densely over fixed ranges,
then each such coordinate contributes a power of $\rho$, because its peak narrows as $1/\rho$
against nodes that do not move:
\begin{equation}
\begin{aligned}
 C_q(\rho)&\ \propto\ \prod_{a\in{\cal D}_q}N_a(\rho)
 \ \propto\ \prod_{a\in{\cal D}_q}\rho^{d_a}\ =\ \rho^{\,d_q},\\
 d_q&\equiv\sum_{a\in{\cal D}_q}d_a ,
\end{aligned}
 \label{eq:quadrature_cost_count}
\end{equation}
where $d_a$ is the effective cost dimension of coordinate $a$: the full dimension where grid
quadrature is required, and less where an adaptive scheme can be exploited.  For the angles, ERAI
at arbitrary density gives $d_a=2$ over $(\phi_{\rm ref},\psi)$; marginalizing
$\psi$ over a local Gaussian peak by Laplace's method leaves $d_a=1$; and treating both angles by
fast local approximations gives $d_a=0$.  Distance and time behave the same way, each contributing
$1$ or $0$ according to the rule adopted.
The coefficients in these laws are Fisher widths, which Appendix~\ref{ap:fisher_coeff} tabulates.
For a representative $35+30\,M_\odot$ binary in the aLIGO
network analysed from $10$ to $1700\,$Hz, the noise-weighted spread is
$\sigma_f=\ftLadderSigF\,$Hz, so the time peak after the angles are marginalized has width
$\sigma_t\rho=\ftLadderTblkMs\,$ms, and the conditional widths of $\ln D$ and $\phi_{\rm ref}$
are $1/\rho$ and $1/(2\rho)$ with no coefficient to measure.
The corresponding errors do not share one exponent.  For a local peak
$I=\int dx\,\exp[A f(x)]$ with $A=\rho^2/2$ and width
$\sigma\propto A^{-1/2}$, their leading magnitudes are
\begin{equation}
 |\Delta\ln I|\sim
 \begin{cases}
   A^{-1}\sim\rho^{-2}, & \text{local shape},\\
   \ln(h/\sigma), & \text{bracketed peak},\\
   Ah^2\sim\rho^2, & \text{missed peak},
 \end{cases}
 \label{eq:quadrature_error_classes}
\end{equation}
where $h$ is the fixed node spacing.  The angular Laplace error belongs to the first class, the
terminal time rule to the second, for which $\ln(h/\sigma)=\ln\rho+O(1)$, and an unresolved
angular product grid to the third.  The uniform distance grid crosses between the last two regimes as the peak narrows, producing a
resolution cliff rather than a useful global error exponent.
Dense
angular rules become more expensive as their nodes follow the narrowing peak, while fixed time
and distance rules eventually fail to resolve it.  The controller instead uses the $U,V,Q$
structure to locate a small set of four-dimensional peaks and accepts their local integral only
when all local acceptance diagnostics pass, including its empirical error budget.
For the time rule the width in the logarithm is the block
width, so $\ln(h/\sigma_t)=\ln(2\pi\sigma_f\rho\,\Delta T)$ is a prediction; for
the representative five-rung sequence it reproduces the measured width to
$\ftLadderRungAgreePct\%$, and only the slope of the logarithm is fitted.
Its width is $\sigma_d=d/\rho$
exactly, so a grid over a fixed prior range needs $N_d\gtrsim\rho\,(D_{\max}-D_{\min})/(c\,d)$
nodes; the representative ladder raises $\rho$ by moving the source closer,
so there $N_d$ grows as $\rho^2$ (Appendix~\ref{ap:fisher_coeff}).
\subsection{Selected configuration}
\label{sec:quadrature:scheme}
Table~\ref{tab:scaling_plan} states the policy as a decision rather than as a menu of independent
algorithms.  A failed local diagnostic retains the likelihood point and selects the band-limited
reserve; the peak-local integral is selected only when every local acceptance diagnostic passes,
including the error-budget check.
The reserve branch of Table~\ref{tab:scaling_plan} rests on the band-limited reconstruction of
$Q$ of Eq.~\eqref{eq:q-time-pregrid}, not on the band-limited terminal quadrature, and so is
admissible wherever the peak-local branch is.
\begin{table}[tbp]\small
\caption{\label{tab:scaling_plan}\textbf{Selected composite marginalization policy.}
The observed rungs refer to the representative amplitude sequence;
selection is made by the conjunction of local acceptance diagnostics, not by a hard SNR threshold.}
\centering
\begin{ruledtabular}
\begin{tabular}{@{}p{0.30\columnwidth} p{0.42\columnwidth} l@{}}
controller disposition & selected treatment & observed rungs \\
\hline
all local acceptance diagnostics pass (including budget) & $U,V,Q$-guided peak-local integral in
  $(t,\phi_{\rm ref},\psi,D)$ & $\rho\simeq\{\mpLLocalRungs\}$ \\
any local acceptance diagnostic declines & band-limited reserve: exact angles, adaptive
  distance, band-limited $Q(t)$ reconstruction & $\rho\simeq\{\mpLReserveRungs\}$ \\
\end{tabular}
\end{ruledtabular}
\end{table}
\emph{The local branch selects and delivers on production data.}  We exercised the four-axis
local integral directly at the two amplitudes where the controller selects it, on
$\aapRungs$ rungs and $\aapSeeds$ seeds.  It was selected on $\aapProdAccept$ of $\aapProdRows$
rows of the likelihood as production builds it, and on $\aapAllAccept$ of $\aapAllRows$ rows
across every configuration tried.  Accepted values lie $\aapErrLo$ to $\aapErrHi\,$nat from an
independent refined reference, while the reserve they replace lies $\aapReserveLo$ to
$\aapReserveHi\,$nat from the same reference: three to four orders of magnitude closer.  A
selected evaluation costs $\aapCostLo$ to $\aapCostHi\,$s at a device peak of $\aapMemLo$ to
$\aapMemHi\,$MiB.  We quote the range rather than a mean because it spans two amplitudes, two
seeds and two stored-window widths, and the spread is the quantity a reader sizing a run needs.
Every declined row failed a budget diagnostic, not an exception.  These numbers are provisional:
they fix the method's accuracy and cost, not the frequency with which the controller selects it,
which a broader injection set is still needed to establish.
In the four-rung test, the controller selects the warranted
reserve at $\rho\simeq\{\mpLReserveRungs\}$ and the peak-local branch at
$\rho\simeq\{\mpLLocalRungs\}$.  The selected result differs from its matched validation reference by at
most $\mpLMaxSelectedError\,$nat.  Once selected, the high-SNR local branch costs about
$\mpLLocalCost\,$s including placement at either rung, so its work follows the number of
located extrema rather than their shrinking widths.  These measurements establish a plausible
operating policy for this configuration; a broader injection set is needed to determine how often
the reserve is selected.
The terminal time rule is the entry this policy leaves open.  Both angular treatments above
marginalize the phase-like angles inside the likelihood, and there the band-limited terminal
quadrature is inadmissible, so the fixed-grid rule is what remains.  Analyses in that regime carry
its error, which by Eq.~\eqref{eq:quadrature_error_classes} grows as $\ln\rho$.  The reserve
branch is unaffected, resting on the reconstruction of $Q$ rather than on a terminal rule.  Closing
the entry requires a rule that interpolates $\ln\mathcal{L}(t)$ and integrates its exponential,
assuming no band limit on the integrand; no released driver offers one.  Fixed angular and uniform-distance grids are excluded
throughout, because their resolution requirement grows with amplitude.
\section{Quadrature over the extrinsic angles, time and distance}
\label{ap:quadrature}
Section~\ref{sec:quadrature} classifies direct marginalization by how its nodes are placed:
fixed-grid rules inherit a resolution cost that grows with amplitude, structure-based rules
reconstruct a known finite representation, and peak-local rules scale their nodes to the integrand.
Here we derive those statements for the reference phase, polarization, arrival time, and luminosity
distance.  In each case the relevant peak has width proportional to \(1/\rho\), but the available
information differs: the angular exponent is a finite trigonometric polynomial, the time-domain
log likelihood is band limited, and the distance exponent is quadratic in inverse distance.
\begin{table*}
\centering
\caption{\label{tab:quadrature_menu}\textbf{Quadrature schemes for directly marginalized
extrinsic parameters.}  The last column gives the amplitude dependence required to maintain
accuracy.  A restricted sampling range leaves the prior density unchanged, so it applies no
evidence correction.}
\begin{ruledtabular}
\begin{tabular}{llll}
axis & scheme & nodes set by & with $\rho$ \\
\hline
$\phi_{\rm ref},\psi$ & product grid          & a fixed count, chosen once            & grows without bound \\
$\phi_{\rm ref},\psi$ & ERAI (code: \texttt{exact}) & mode content and amplitude & calls constant; arithmetic $\rho^2$ \\
$\psi$                & analytic Laplace      & --- (no quadrature)                   & error falls as $1/A$ \\
$\psi$                & peak-local            & stationary points, exactly            & constant (4 cells) \\
$\phi_{\rm ref},\psi$ & joint analytic Laplace & --- (no quadrature)                  & error falls as $1/A$ \\
$t$                   & fixed-spacing Simpson & the data sample rate                  & requirement grows linearly \\
$t$                   & local interpolation of stored $Q$ samples & the interpolation order       & error grows as $\rho^2$ \\
$t$                   & full-window band-limited interpolation & a factor derived from the peak width & grows as the peak sharpens \\
$t$                   & peak-local            & a neighbourhood of the located peak   & constant \\
$d$                   & uniform grid            & the prior range, count fixed        & error grows without bound \\
$d$                   & log-uniform, tolerance-sized & a stated tolerance and the amplitude & count grows; error held \\
$d$                   & adaptive Gauss--Hermite & the closed-form peak and width      & constant \\
$d$                   & restricted sampling range & a located peak; prior unchanged   & constant \\
sky                   & restricted sampling range & an external localization; prior unchanged & constant \\
\end{tabular}
\end{ruledtabular}
\end{table*}
\subsection{Angles: finite harmonic content}
At fixed arrival time, the complex antenna factor contributes \(e^{2i\psi}\), while a waveform
mode \(m\) contributes \(e^{im\phi_{\rm ref}}\).  The data term in
Eq.~\eqref{eq:lnL:MatrixForm} is linear in these factors and the model norm is quadratic.
Consequently,
\begin{equation}
\ln\mathcal{L}(\phi_{\rm ref},\psi\,|\,t,\bm\lambda)=
\sum_{p=-2}^{2}\ \sum_{\substack{q=-2m_{\max}\\ q\ \mathrm{even}}}^{2m_{\max}}
c_{pq}(t,\bm\lambda)e^{i(2p\psi+q\phi_{\rm ref})}.
\label{eq:ap:angleharm}
\end{equation}
The polarization series terminates at \(4\psi\), independently of waveform content.  The
reference-phase series terminates at \(2m_{\max}\); choosing
\(m_{\max}=\ell_{\max}\) also covers precessing models, because rotations mix the \(m\) components
within a fixed \(\ell\) without creating higher \(\ell\).  A product grid satisfying
\begin{equation}
N_\psi\ge5,\qquad N_{\phi_{\rm ref}}\ge2m_{\max}+1
\label{eq:ap:nyquist}
\end{equation}
therefore determines every coefficient.  ERAI uses this exact finite model, while sizing the
subsequent numerical integral of \(e^{\ln\mathcal L}\) from the reconstructed amplitude.  Thus the
expensive likelihood count is fixed by waveform mode content rather than by \(\rho\), although the
inexpensive arithmetic grid still refines with amplitude.  The implementation currently exposes
ERAI under the option name \texttt{exact}; the name refers only to reconstruction, not to a
closed-form marginal integral.
The fixed-time qualification is essential.  Equation~\eqref{eq:ap:angleharm} must be reconstructed
at each time sample before time marginalization.  Integrating over time first applies a nonlinear
log-sum-exp operation, and the resulting angular function need not retain the finite harmonic
support of the fixed-time exponent.
The same structure supplies two one-dimensional treatments of polarization.  With \(u=2\psi\) at
fixed \(\phi_{\rm ref}\),
\begin{equation}
q(u)=a+\mathrm{Re}\,c_{1}e^{iu}+\mathrm{Re}\,c_{2}e^{2iu}.
\label{eq:ap:psiexp}
\end{equation}
If \(q\) has one dominant maximum \(u_\star\), with
\(A=-q''(u_\star)>0\), Laplace's method gives
\begin{equation}
\int du\,e^{q(u)}
=e^{q(u_\star)}\sqrt{\frac{2\pi}{A}}
\Bigl[1+O\!\left(A^{-1}\right)\Bigr].
\label{eq:ap:psilaplace}
\end{equation}
Since \(A\propto\rho^2\), the relative error decreases as \(O(\rho^{-2})\).  This prediction is
also observed in the amplitude ladder: for the \(\vaLAngErrNRungs\) tested amplitudes,
the product of error and \(A=\vaLAngErrTimesADef\) is
\(\vaLAngErrTimesA\) nats to within \(\vaLAngErrTimesAScatter\%\) across a factor
\(\vaLAngErrASpan\) in \(A\).
When a single-mode Laplace approximation is not appropriate, differentiating
Eq.~\eqref{eq:ap:psiexp} and setting \(z=e^{iu}\) gives a quartic equation.  Its unit-circle roots
enumerate all stationary points; sorting them partitions the circle into disjoint cells \(I_m\):
\begin{equation}
\int_{0}^{2\pi}du\,e^{q(u)}
=\sum_m\int_{I_m}du\,e^{q(u)}.
\label{eq:ap:psicells}
\end{equation}
Each cell is integrated at a density set by its own curvature.  Completeness follows from the
quartic enumeration; numerical refinement is then confined to the integrals within the cells.
\subsection{Time: representing the correlations and placing the integration nodes}
Time marginalization involves three distinct operations: storing each detector--mode correlation
\(Q_{k,\ell m}(t)\), evaluating it at detector arrival times between stored samples, and integrating
the resulting likelihood over geocentric coalescence time.  The first two determine the numerical
representation of \(Q\); the third determines the quadrature nodes.  Resolving one operation does
not guarantee that the other is accurate.
The frequency-domain definition of \(Q_{k,\ell m}\) contains no frequencies above the waveform
cutoff, so \(Q\) is band limited.  Let \(h_Q\) be the spacing of its stored samples and
\(f_Q=1/h_Q\) their sampling rate.  The sampling theorem then gives the continuous correlation as
\begin{equation}
Q_{k,\ell m}(t)=
\sum_j Q_{k,\ell m}(t_j)\,
\mathrm{sinc}\!\left(\frac{t-t_j}{h_Q}\right),
\label{eq:ap:sinc}
\end{equation}
where \(\mathrm{sinc}(x)=\sin(\pi x)/(\pi x)\).  Repeatedly evaluating a long sinc stencil at every
requested detector time is expensive, whereas truncating the stencil produced measurable errors
in our high-amplitude tests.  We instead build a fine table once, when the correlation bank is
packed.  We append a reversed copy of the finite series so that its periodic Fourier extension is
continuous, zero-pad its Fourier coefficients, and transform back to obtain
\begin{equation}
\begin{aligned}
 Q^{(r)}&=\mathcal F^{-1}\!\left[\mathcal Z_r\mathcal F(Q_{\rm refl})\right],
 &\delta t_Q&=h_Q/r,\\
 \widehat Q(t)&=\mathcal C_3[Q^{(r)}](t).
\end{aligned}
 \label{eq:q-time-pregrid}
\end{equation}
where \(Q_{\rm refl}\) is the reflected series, \(\mathcal Z_r\) pads its Fourier coefficients to
produce spacing \(\delta t_Q\), and \(\mathcal C_3\) is a four-point cubic interpolation for the
remaining sub-grid displacement.  The Fourier step is paid once; the much cheaper cubic is paid
inside each likelihood evaluation.  Thus the cubic does not replace band-limited reconstruction:
it evaluates an already reconstructed fine grid without repeating a sinc sum for every detector
delay.
For the preferred conventional and differentiable amplitude-ladder calculations in this paper,
\(f_Q=\toaNativeQRateHz\,\mathrm{Hz}\), corresponding to a Nyquist frequency of
\(\toaQNativeNyquistHz\,\mathrm{Hz}\); the waveform correlations end at
\(\toaQBandMaxHz\,\mathrm{Hz}\).  We use the single value \(r=\toaPregridFactor\), so the refined
table has the effective sampling rate and spacing
\begin{equation}
 f_Q^{(r)}=r f_Q=\toaPregridRateHz\,\mathrm{Hz},
 \qquad
 \delta t_Q=\frac{1}{f_Q^{(r)}}=\toaPregridSpacingMicroseconds\,\mu\mathrm{s}.
 \label{eq:ap:q-pregrid-rate}
\end{equation}
Nyquist sampling guarantees that the native table contains the waveform bandwidth, but not that
interpolation from that table meets an inference-level accuracy target.  For a phase-marginalized
quadrupole signal, the characteristic correlation timescale is \((2\pi\sigma_f)^{-1}\), where
\(\sigma_f\) is the noise-weighted frequency spread of Eq.~\eqref{eq:ap:fmoments}.  The leading
four-point cubic error scales as
\begin{equation}
 \begin{aligned}
 |\Delta\ln\mathcal L|&\ \propto\
 \rho^2\bigl(2\pi\sigma_f\delta t_Q\bigr)^4,\\
 r_{\rm required}&\ \propto\ \sigma_f\sqrt{\rho}.
 \end{aligned}
 \label{eq:ap:q-pregrid-accuracy}
\end{equation}
At fixed error budget and native rate, this scaling supplies an initial value of \(r\), not an
acceptance test, because waveform content and detector delays determine its coefficient.  In
practice we choose the smallest refinement that satisfies declared pointwise
\(|\Delta\ln\mathcal L|\) and evidence \(|\Delta\ln Z|\) tolerances against a denser Fourier
reference; doubling \(r\) should reduce the cubic contribution by approximately a factor of
sixteen until another error dominates.
For the calculations reported here, \(r=\toaPregridFactor\) passes the predeclared
\(\toaAccuracyBudget\)-nat bounds for both quantities across the validation matrix, including its
loudest endpoint at \(\rho=\toaValidationSnrMax\).  The largest measured errors are
\(\toaCandidateWorstLnL\) nat pointwise and \(\toaCandidateWorstLnZ\) nat in evidence
(Table~\ref{tab:q-interpolation-oracle}).  We use this qualified value at every ladder rung as a
conservative fixed setting.  It is not universal: sources with larger \(\rho\sigma_f\), different
frequency support, or a different native rate require a new oracle check.
The refined correlation rate is not the sampling rate of the final likelihood integral.  At the
loudest ladder point, the likelihood peak is only
\(\sigma_t\simeq\ftLadderSigmaTHiUs\,\mu\mathrm{s}\) wide, so the integration rule described
below must request enough arrival times to resolve it.
\begin{table*}[tbp]
\caption{\label{tab:q-interpolation-oracle}\textbf{Detector-correlation interpolation against a
direct Fourier reference.}  The columns report the largest absolute pointwise log-likelihood
difference and the posterior-reweighted evidence correction, both in nats.  Each row contains
100 fixed posterior draws.  The acceptance bound is \(\toaAccuracyBudget\) nat for each reported
quantity.}
\centering\small
\begin{tabular}{llrrrr}
\hline
Model & SNR & \multicolumn{2}{c}{Finite-sinc interpolation} & \multicolumn{2}{c}{32-kHz table + four-point cubic} \\
 & & max $|\Delta\ln L|$ & $\Delta\ln Z$ & max $|\Delta\ln L|$ & $\Delta\ln Z$ \\
\hline
22 & 40.77 & 0.126 & 0.0615 & 2.16e-05 & 1.06e-05 \\
22 & 652.31 & 17 & 16.5 & 0.00353 & -0.00121 \\
HM & 51.38 & 0.0688 & 0.0257 & 9.97e-05 & 1.35e-05 \\
HM & 72.11 & 0.118 & 0.0595 & 0.000133 & 1.82e-05 \\
\hline
\end{tabular}

\end{table*}
At fixed values of the other extrinsic parameters, the weighted correlation combination entering
the data term is also band limited.  Its exponential is not: if
\(\ell(t)=\ln\mathcal L(t)\), the fixed-spacing rule
\begin{equation}
\int dt\,e^{\ell(t)}
\simeq h\sum_k w_k e^{\ell(t_k)},
\qquad h=1/f_{\rm s},
\label{eq:ap:tfixed}
\end{equation}
where \(w_k\) are the integration weights and \(f_{\rm s}\) is the geocentric-time integration
rate.  After angular marginalization the exponential peak has approximate width
\begin{equation}
\sigma_t=\frac{1}{2\pi\rho\,\sigma_f},
\label{eq:ap:sigmat}
\end{equation}
so requiring \(h\lesssim\sigma_t\) implies
\begin{equation}
f_{\rm s}\gtrsim2\pi\sigma_f\rho.
\label{eq:ap:srate}
\end{equation}
Thus a data rate sufficient to represent the correlation series need not be a sufficient
quadrature rate, and the latter grows linearly with amplitude.
The \emph{band-limited full-window rule} reconstructs the weighted correlation combination
\(\kappa(t)\) on a finer geocentric-time grid and only then evaluates the likelihood.  If \(M\) fine
intervals are placed within each original interval \(h\),
\begin{equation}
\begin{aligned}
\int dt\,e^{\ell(t)}
&\simeq\frac{h}{M}\sum_k\exp\!\left[\ell\!\left(
\sum_j\kappa(t_j)
\mathrm{sinc}\!\left(\frac{t_k-t_j}{h}\right)
\right)\right],\\
M&\gtrsim h/\sigma_t ,
\end{aligned}
\label{eq:ap:tband}
\end{equation}
where \(\ell(\kappa)\) is evaluated with the other extrinsic parameters fixed.  This requires no
new waveform--data inner products, but its arithmetic grows as \(M\propto\rho\).  The order matters:
after phase, polarization, or distance has been marginalized at each time, nonlinear sums destroy
the finite-band representation.  Equation~\eqref{eq:ap:tband} therefore acts on \(\kappa\), not on
an already marginalized likelihood.  Interpolation of \(Q\) remains valid because \(Q\) itself is
band limited.
The \emph{peak-local rule} instead locates every maximum of the reconstructed \(\kappa(t)\) and
integrates only curvature-sized neighborhoods, merging overlaps:
\begin{equation}
\begin{aligned}
\int dt\,e^{\ln\mathcal{L}(t)}
&=\sum_m\int_{I_m}dt\,e^{\ln\mathcal{L}(t)}+\mathcal R,\\
I_m&=[\hat t_m-c\sigma_t,\hat t_m+c\sigma_t].
\end{aligned}
\label{eq:ap:tlocal}
\end{equation}
If \(\Omega\) is the complete time window, the omitted contribution obeys
\(0\le\mathcal R\le
|\Omega\setminus\cup_m I_m|\exp[\sup_{\Omega\setminus\cup_m I_m}\ln\mathcal L]\).
If this bound exceeds the requested tolerance, or the local calculation would cost more than the
full-window rule, we use Eq.~\eqref{eq:ap:tband}.  We also decline the local rule when prior phase
marginalization prevents a justified width estimate.
\subsection{Distance: locating a moving peak}
Let \(x=d_{\rm ref}/d\).  Because the waveform amplitude scales as \(1/d\), the two terms in
Eq.~\eqref{eq:def:lnL:Decomposed} are exactly linear and quadratic in \(x\):
\begin{equation}
\ln\mathcal{L}(x)=Kx-\tfrac12Rx^2,
\label{eq:ap:dexp}
\end{equation}
where \(K\) is the data--template inner product and \(R=\langle h|h\rangle\), both evaluated using
the reference-distance waveform.  Completing the square gives
\begin{equation}
x_\star=K/R,\qquad \sigma_x=R^{-1/2},
\label{eq:ap:dpeak}
\end{equation}
and \(\sigma_x/x_\star\sim1/\rho\).  For a prior uniform in Euclidean volume, the transformed
density is \(p_x(x)\propto x^{-4}\); it modulates but does not change the Gaussian localization of
the likelihood exponent.
A grid uniform in \(d\) spans the prior independently of the peak:
\begin{equation}
\mathcal{L}_d\simeq
\sum_jw_jp(d_j)e^{\ln\mathcal{L}(d_j)},
\qquad \Delta d\ \text{fixed by the prior range}.
\label{eq:ap:dgrid}
\end{equation}
Its spacing must shrink with the peak width, so the node count required for fixed accuracy grows
with \(\rho\).  In contrast, an adaptive Gauss--Hermite rule centers and scales its fixed
abscissae using Eq.~\eqref{eq:ap:dpeak}:
\begin{equation}
\mathcal{L}_d\simeq
\sum_i\widetilde w_i\,
p_x(x_\star+\xi_i\sigma_x)
e^{\ln\mathcal{L}(x_\star+\xi_i\sigma_x)} .
\label{eq:ap:dgh}
\end{equation}
All amplitude dependence is absorbed into \(x_\star\) and \(\sigma_x\), leaving a node count set by
the shape in local-width units.  This also handles the distance--inclination degeneracy: the peak
can move away from a fiducial distance while narrowing, so merely shrinking a fixed interval about
that fiducial point is not reliable.
Restricting the sampled interval \(\mathcal B\) is distinct from changing the prior.  If the prior
density is left unrenormalized,
\begin{equation}
\mathcal{L}
=\underbrace{\int_{\mathcal B}dd\,p(d)e^{\ln\mathcal L(d)}}_
{\mathcal L_{\mathcal B}}
+\int_{\mathcal B^c}dd\,p(d)e^{\ln\mathcal L(d)} ,
\label{eq:ap:dbox}
\end{equation}
so \(\mathcal L_{\mathcal B}\le\mathcal L\), and the omitted term is precisely the probability
mass that must be bounded.  Stability under a modest widening is insufficient if the peak lies
outside both intervals; the analytic \(x_\star\) provides the necessary location check.  The same
logic applies to a sky box supplied by an external localization.
A log-uniform grid offers a tolerance-controlled fixed-grid alternative.  For a Gaussian peak with
relative width \(s\), Poisson summation bounds the leading alias by
\(2\exp(-2\pi^2s^2/h^2)\).  Setting this term to a requested tolerance and using
\(s\simeq1/\rho_{\max}\) gives
\begin{equation}
\Delta(\ln d)\le\frac{c(\mathrm{tol})}{\rho_{\max}},
\qquad
c(\mathrm{tol})=\pi\sqrt{\frac{2}{\ln(2/\mathrm{tol})}} .
\label{eq:ap:dlog}
\end{equation}
The count grows with the declared amplitude bound so that the discretization error remains
controlled.  The rule must also reject a peak outside the interval or too near an endpoint, where
the infinite-grid alias estimate alone does not bound the omitted tail.
These derivations supply the coordinate-specific refinements used in
Eq.~\eqref{eq:direct_marginalization_validation}.  To test a production configuration, one
coordinate is refined at a time while the remaining quadratures and sample locations are fixed.
The resulting \(\Delta_a\) bounds discretization on that coordinate; Eqs.~\eqref{eq:ap:tlocal}
and~\eqref{eq:ap:dbox} show why omitted support must be checked separately.  Agreement of the total
marginal alone cannot replace these axis-wise tests because signed errors from different
coordinates may cancel.

\section{Fisher coefficients for the resolution rules}
\label{ap:fisher_coeff}
The scaling laws of Section~\ref{sec:quadrature} and Appendix~\ref{ap:quadrature} carry
exponents in $\rho$ but no coefficients.  The coefficients are Fisher widths, and for a static
detector they follow from noise-weighted moments of the template
\cite{CutlerFlanagan:1994,1995PhRvD..52..848P,2009NJPh...11l3006F}.  This appendix tabulates them
for 2G and 3G noise curves, checks them against GWFish~\cite{2023A&C....4200671D}, and states the
rules with their coefficients.  The detector is held at its orientation and position at the merger
time, and the long-wavelength response is used.  GWFish with Earth rotation on changes the sky area
of the longest signal below (a BNS from $\ftRotationStartHz\,$Hz in CE$+$ET$+$K) by
$\ftRotationAreaPct\%$, and the arm transfer function changes no area by more than
$\ftFiniteArmAreaPct\%$, so neither assumption moves a coefficient.
\subsection{Widths from noise moments}
The Fisher matrix $\Gamma_{ab}=\langle\partial_a h|\partial_b h\rangle$ scales as $\rho^2$, so every
width below is quoted as $\sigma\rho$ and every area as $A_{90}\rho^2$.  For one detector and one
polarization mode, $h(f)=A\,e^{2i\phi}e^{2\pi ift}\tilde h(f)$, and the derivatives with respect to
$\ln A$, $\phi$, and $t$ are $h$, $2ih$, and $2\pi ifh$.  With the moments
\begin{equation}
\langle f^n\rangle=\frac{4}{\rho^2}\int\!df\,\frac{f^n|h(f)|^2}{S_n(f)},\qquad
\sigma_f^2=\langle f^2\rangle-\langle f\rangle^2,
\label{eq:ap:fmoments}
\end{equation}
the block is
\begin{equation}
\Gamma=\rho^2\begin{pmatrix}1&0&0\\0&4&4\pi\langle f\rangle\\0&4\pi\langle f\rangle&4\pi^2\langle f^2\rangle\end{pmatrix}
\label{eq:ap:fisher3}
\end{equation}
in the order $(\ln A,\phi,t)$.  Amplitude is in quadrature with the phase-like parameters, so
$\sigma_{\ln A}\rho=1$ at every level of conditioning.  Holding $\phi$ fixed gives
$\sigma_t\rho=1/(2\pi f_{\rm rms})$ with $f_{\rm rms}=\langle f^2\rangle^{1/2}$; marginalizing
$\phi$ gives Eq.~\eqref{eq:ap:sigmat}, $\sigma_t\rho=1/(2\pi\sigma_f)$.  The phase width is
$\sigma_\phi\rho=1/2$ conditional and $f_{\rm rms}/(2\sigma_f)$ marginal.  Freeing the masses
widens $\sigma_t$ again, because the chirp-mass phase derivative is correlated with time
\cite{1995PhRvD..52..848P}.
Three levels of conditioning occur in the pipeline, and they answer different questions:
\begin{itemize}
\item \emph{conditional}: one axis with every other parameter fixed.  The width of a single
  quadrature axis at fixed nodes on the others.
\item \emph{block}: the four in-likelihood axes $(\ln D,\psi,\phi,t)$ free, sky and inclination
  fixed at the sampled values.  The time peak that remains after the angles are marginalized, which
  the terminal time rule must resolve.
\item \emph{marginal}: all seven extrinsic parameters free, or nine with $(\ln\mathcal{M}_c,\eta)$.
  The width the outer sampler sees.
\end{itemize}
Standard Fisher tables quote the marginal level.  The quadrature rules need the first two.
For a network the sky enters through the antenna factors $F_k(\alpha,\delta,\psi)$ and the delays
$\tau_k(\alpha,\delta)$, and the seven-parameter matrix is assembled from the same inner products
with the sky derivatives taken numerically.  The 90\% area is
$A_{90}=\pi\chi^2_{0.9}(2)\sqrt{\det C_{\rm sky}}$ from the marginal covariance $C$.  The
timing-only triangulation of \cite{2009NJPh...11l3006F} uses per-detector
$\sigma_{t,k}=1/(2\pi\rho_k\sigma_{f,k})$ and gives an upper bound on the area; the full matrix
adds amplitude and phase consistency across sites.
\subsection{Coefficients}
Table~\ref{tab:fisher_moments} lists the moments and single-detector widths for four generic
sources, for the loud binary-neutron-star injection of Section~\ref{sec:3g_demo}, and for a $35+30\,M_\odot$ binary in aLIGO from $10$ to $1700\,$Hz, and
Table~\ref{tab:fisher_network} the marginal widths and sky areas for the three networks used in
this paper.  The 2G rows start at $\ftTwoGFminHz\,$Hz and the 3G rows at
$\ftRotationStartHz\,$Hz, both to $\ftFmaxHz\,$Hz; the loud injection uses its own band of
$\ftEventBFminHz$ to $\ftEventBFmaxHz\,$Hz.  Its sky position and orientation are used throughout.
The semi-analytic values reproduce the closed forms of Eq.~\eqref{eq:ap:fisher3} to $10^{-4}$, and
change by at most $\ftGridConvergencePct\%$ when the frequency spacing is quartered.
\begin{table*}[tbp]
\caption{\label{tab:fisher_moments}\textbf{Noise-weighted moments and single-detector time widths.}
Widths are $\sigma_t\rho$ in ms at three levels: everything else fixed
[$1/(2\pi f_{\rm rms})$], amplitude and phase free [$1/(2\pi\sigma_f)$], and chirp mass and
symmetric mass ratio also free.  The last column is the fixed-grid sample rate per unit $\rho$,
$f_{\rm s}/\rho=2\pi\sigma_f$, from Eq.~\eqref{eq:ap:srate}.}
\centering\small
\begin{ruledtabular}
\begin{tabular}{llrrrrrrrr}
source & PSD & $f_{\min}$ [Hz] & $\langle f\rangle$ [Hz] & $f_{\rm rms}$ [Hz] & $\sigma_f$ [Hz] &
$\sigma_t\rho$ cond.\ [ms] & $\sigma_t\rho$ block [ms] & $\sigma_t\rho$ $+$masses [ms] & $f_{\rm s}/\rho$ [Hz] \\
\hline
BNS $1.4+1.4$ & aLIGO & $20$ & $100$ & $150$ & $120$ & $1.0$ & $1.4$ & $2.6$ & $730$ \\
BNS $1.4+1.4$ & A$+$ & $20$ & $120$ & $180$ & $130$ & $0.88$ & $1.2$ & $2.3$ & $840$ \\
BNS $1.4+1.4$ & CE & $5$ & $43$ & $71$ & $57$ & $2.2$ & $2.8$ & $4.7$ & $360$ \\
BNS $1.4+1.4$ & ET & $5$ & $55$ & $110$ & $93$ & $1.5$ & $1.7$ & $2.7$ & $590$ \\
NSBH $10+1.4$ & aLIGO & $20$ & $85$ & $120$ & $86$ & $1.3$ & $1.8$ & $3.2$ & $540$ \\
NSBH $10+1.4$ & A$+$ & $20$ & $99$ & $140$ & $100$ & $1.1$ & $1.6$ & $2.7$ & $640$ \\
NSBH $10+1.4$ & CE & $5$ & $38$ & $58$ & $44$ & $2.7$ & $3.6$ & $6.3$ & $280$ \\
NSBH $10+1.4$ & ET & $5$ & $43$ & $82$ & $70$ & $1.9$ & $2.3$ & $3.6$ & $440$ \\
BBH $30+30$ & aLIGO & $20$ & $91$ & $110$ & $70$ & $1.4$ & $2.3$ & $9.0$ & $440$ \\
BBH $30+30$ & A$+$ & $20$ & $110$ & $130$ & $77$ & $1.2$ & $2.1$ & $7.9$ & $480$ \\
BBH $30+30$ & CE & $5$ & $39$ & $59$ & $44$ & $2.7$ & $3.6$ & $7.0$ & $280$ \\
BBH $30+30$ & ET & $5$ & $45$ & $78$ & $64$ & $2.0$ & $2.5$ & $4.9$ & $400$ \\
BBH $80+80$ & aLIGO & $20$ & $65$ & $70$ & $26$ & $2.3$ & $6.2$ & $16$ & $160$ \\
BBH $80+80$ & A$+$ & $20$ & $68$ & $73$ & $27$ & $2.2$ & $6.0$ & $14$ & $170$ \\
BBH $80+80$ & CE & $5$ & $36$ & $44$ & $26$ & $3.6$ & $6.1$ & $24$ & $160$ \\
BBH $80+80$ & ET & $5$ & $31$ & $45$ & $32$ & $3.6$ & $5.0$ & $17$ & $200$ \\
loud BNS $1.6+1.4$ & CE & $50$ & $110$ & $140$ & $86$ & $1.1$ & $1.9$ & $5.8$ & $540$ \\
loud BNS $1.6+1.4$ & ET & $50$ & $140$ & $190$ & $120$ & $0.86$ & $1.4$ & $4.2$ & $730$ \\
Fig.~2 ladder $35+30$ & aLIGO & $10$ & $86$ & $110$ & $66$ & $1.5$ & $2.4$ & $7.1$ & $410$ \\
Fig.~2 ladder $35+30$ & A$+$ & $10$ & $99$ & $120$ & $72$ & $1.3$ & $2.2$ & $7.0$ & $450$ \\

\end{tabular}
\end{ruledtabular}
\end{table*}
\begin{table*}[tbp]
\caption{\label{tab:fisher_network}\textbf{Network widths for static detectors at the loud-injection sky
position.}  $\sigma_t\rho$ in ms at the block level and with all seven extrinsic parameters free;
the remaining widths are marginal, in radians for the angles.  $A_{90}\rho^2$ in deg$^2$ with all
seven free and with the arrival time held fixed.  The last column is the ratio of the GWFish area
to the semi-analytic one, with the delay sign made consistent with its waveform convention.}
\centering\small
\begin{ruledtabular}
\begin{tabular}{llrrrrrrrrr}
source & network & $\sigma_t\rho$ block & $\sigma_t\rho$ marg. & $\sigma_{\ln D}\rho$ &
$\sigma_\iota\rho$ & $\sigma_\psi\rho$ & $\sigma_\phi\rho$ & $A_{90}\rho^2$ & $A_{90}\rho^2$, $t$ fixed & GWFish \\
\hline
BNS $1.4+1.4$ & HLV & $1.4$ & $9.8$ & $5.6$ & $6.6$ & $6.4$ & $5.4$ & $6800$ & $950$ & $1.00$ \\
BNS $1.4+1.4$ & CE$+$ET & $2.3$ & $27$ & $4.3$ & $3.3$ & $4.1$ & $4.1$ & $3.9\times10^{4}$ & $3400$ & $0.99$ \\
BNS $1.4+1.4$ & CE$+$ET$+$K & $2.4$ & $9.8$ & $3.6$ & $2.7$ & $3.2$ & $3.4$ & $8600$ & $2100$ & $1.01$ \\
NSBH $10+1.4$ & HLV & $1.8$ & $12$ & $5.7$ & $6.7$ & $6.6$ & $5.5$ & $1.0\times10^{4}$ & $1600$ & $1.00$ \\
NSBH $10+1.4$ & CE$+$ET & $3.1$ & $27$ & $4.4$ & $3.3$ & $4.1$ & $4.1$ & $4.7\times10^{4}$ & $5400$ & $0.99$ \\
NSBH $10+1.4$ & CE$+$ET$+$K & $3.1$ & $12$ & $3.6$ & $2.7$ & $3.3$ & $3.4$ & $1.2\times10^{4}$ & $3200$ & $1.01$ \\
BBH $30+30$ & HLV & $2.3$ & $13$ & $6.2$ & $7.1$ & $6.6$ & $5.5$ & $1.2\times10^{4}$ & $2000$ & $1.00$ \\
BBH $30+30$ & CE$+$ET & $3.2$ & $27$ & $4.3$ & $3.3$ & $4.1$ & $4.1$ & $4.7\times10^{4}$ & $5600$ & $0.99$ \\
BBH $30+30$ & CE$+$ET$+$K & $3.2$ & $12$ & $3.6$ & $2.7$ & $3.2$ & $3.4$ & $1.2\times10^{4}$ & $3300$ & $1.01$ \\
BBH $80+80$ & HLV & $6.2$ & $18$ & $6.2$ & $7.0$ & $7.0$ & $5.8$ & $2.1\times10^{4}$ & $7200$ & $1.00$ \\
BBH $80+80$ & CE$+$ET & $5.7$ & $27$ & $4.4$ & $3.4$ & $4.1$ & $4.1$ & $6.4\times10^{4}$ & $1.4\times10^{4}$ & $0.99$ \\
BBH $80+80$ & CE$+$ET$+$K & $5.8$ & $16$ & $3.5$ & $2.7$ & $3.3$ & $3.4$ & $2.0\times10^{4}$ & $7200$ & $1.00$ \\
loud BNS $1.6+1.4$ & CE$+$ET & $1.6$ & $25$ & $4.0$ & $3.1$ & $3.7$ & $3.8$ & $2.1\times10^{4}$ & $1400$ & $0.99$ \\
loud BNS $1.6+1.4$ & CE$+$ET$+$K & $1.6$ & $6.5$ & $3.5$ & $2.7$ & $3.1$ & $3.3$ & $3300$ & $830$ & $1.02$ \\
Fig.~2 ladder $35+30$ & HLV & $2.4$ & $14$ & $6.1$ & $7.1$ & $6.7$ & $5.5$ & $1.2\times10^{4}$ & $2200$ & $1.00$ \\

\end{tabular}
\end{ruledtabular}
\end{table*}
The ladder row is tested directly.  With the $\Delta T=\ftLadderDeltaTUs\,\mu$s grid of that
figure, the block width gives $\Delta T/\sigma_t=\ftLadderTkDerived\,\rho$ against
$\ftLadderTkFitted\,\rho$ fitted to the widths measured at the five rungs, and no rung differs
by more than $\ftLadderRungAgreePct\%$.
\subsection{The rules with coefficients}
\begin{enumerate}
\item \emph{Fixed-grid time quadrature.}  Eq.~\eqref{eq:ap:srate} with the block width reads
  $f_{\rm s}\gtrsim\rho/(\sigma_t\rho)$.  For the loud injection this is $\ftFsNetEventBCEETK$ times $\rho$,
  so $\ftEventBFsKHzAtThousand\,$kHz at $\rho=1000$, against the $\jsrRateHz\,$Hz
  configuration used for the loud injection.  The requirement is set by the template, so a
  contemporary source in aLIGO has the same coefficient to within a factor of two
  (Table~\ref{tab:fisher_moments}).
\item \emph{Sub-sample timing error.}  A peak displaced by $\delta$ from where the reconstruction
  places it costs $\Delta\ln\mathcal{L}=(\delta/\sigma_t)^2/2$, so an error budget of
  $\epsilon$ nat allows $\delta\le\sigma_t\sqrt{2\epsilon}$, which is $\ftBudgetFactor\,\sigma_t$
  for a $10^{-3}$-nat error budget.  For the loud injection at $\rho=1000$
  the block width is $\ftEventBSigmaTUsAtThousand\,\mu$s and the allowed error
  $\ftEventBSubsampleBudgetNs\,$ns.  This bound applies to the reconstruction of $Q$ in
  Eq.~\eqref{eq:q-time-pregrid} whatever terminal rule follows it.
\item \emph{Distance.}  The conditional width is $\sigma_{\ln D}\rho=1$ with no coefficient to
  measure.  A grid uniform in $D$ over $[D_{\min},D_{\max}]$ needs
  $N_D\gtrsim\rho\,(D_{\max}-D_{\min})/(c\,D)$ nodes, and a log-uniform grid
  $N_D\gtrsim\rho\ln(D_{\max}/D_{\min})/c$, which is $\ftLogDistanceRange\,\rho/c$ for the prior of
  Table~\ref{tab:quadrature_menu}; the adaptive rule of Eq.~\eqref{eq:ap:dgh} has no $\rho$
  dependence. In the representative amplitude ladder the source moves from
  $\ftLadderDistLo$ to $\ftLadderDistHi\,$Mpc, so $\sigma_d$ falls as $\rho^{-2}$ and the three
  evaluated points lie at $h/\sigma_d=\ftLadderHsigLo$, $\ftLadderHsigMid$, and $\ftLadderHsigHi$.
  The refinement series at $\rho\simeq\ftLadderRhoMid$ passes from $\ftLadderErrRefineA$ to
  $\ftLadderErrRefineB\,$nat between $h/\sigma_d=\ftLadderHsigRefineA$ and
  $\ftLadderHsigRefineB$, so $c\simeq\ftLadderHsigThreshold$ for the $10^{-3}$-nat budget and the
  uniform grid needs $\ftLadderNdLo$, $\ftLadderNdMid$, and $\ftLadderNdHi$ nodes at
  $\rho=\ftLadderRhoLo$, $\ftLadderRhoMid$, and $\ftLadderRhoHi$.  The marginal width the outer
  sampler sees is $\sigma_{\ln D}\rho\simeq\ftLnDmargEventBCEETK$ for the loud injection
  (Table~\ref{tab:fisher_network}), set by the inclination degeneracy.
\item \emph{Angles.}  ERAI's finite-model reconstruction in Eq.~\eqref{eq:ap:nyquist} needs no
  Fisher-width coefficient; its subsequent arithmetic integral is amplitude sized.
  A Laplace treatment needs the conditional widths, $\sigma_\phi\rho=1/2$ and
  $\sigma_\psi\rho\simeq\ftPsiCondEventBCEETK$, to be small against the period of the
  trigonometric polynomial; in the representative amplitude ladder, the Laplace treatment fails below
  $\rho\simeq5$ where $\sigma_\psi\rho\simeq\ftLadderPsiCond$ stops being
  small against the period.  The marginal widths are several radians per unit $\rho$
  (Table~\ref{tab:fisher_network}), so below $\rho\sim10$ the marginal angles wrap and the local
  form does not apply to the outer sampler.
\item \emph{Sky and time for the outer sampler.}  $A_{90}\rho^2$ is $\ftAreaEventBCEETK\,$deg$^2$
  for the loud injection in CE$+$ET$+$K, so the fraction of a broad sky proposal that lands in the basin falls
  as $\rho^{-2}$ with that coefficient.  The sky-marginalized time width is
  $\ftEventBTmargOverBlk$ times the block width, because sky position and arrival time are
  degenerate; a time proposal in the outer sampler must be that wide while the in-likelihood window
  need only cover the block width.
\end{enumerate}
Holding the arrival time fixed shrinks the loud-injection area by the factor $\ftEventBTfixRatio$
(Table~\ref{tab:fisher_network}).  A fixed-grid time rule whose spacing exceeds $\sigma_t$
integrates only the node nearest the peak, which is the time-fixed likelihood.  The two loud-injection
areas this paper reports, $A_{90}\rho^2\simeq\ftPaperAreaFixedRhoSq\,$deg$^2$ under the
fixed-grid rule (Section~\ref{sec:3g_demo}) and $\ftPaperAreaBandRhoSq\,$deg$^2$ under a
band-limited rule, stand in the ratio $\ftPaperTfixRatioMeasured$.
The static long-wavelength area is within $\ftEventBAreaVsPaperBandPct\%$ of the latter.  A
sampler and a curvature check that share the likelihood agree with each other in either case, so
the agreement in Section~\ref{sec:3g_demo} does not by itself certify the time rule.
\subsection{GWFish comparison}
GWFish~$\ftGwfishVersion$ was run on the same sources with the same PSDs written to its detector
files, the same sites and arm azimuths, the same waveform, and its projection frozen at the merger
time.  The conditional and block time widths agree to $\ftGwCondAgreePct\%$.  As shipped, its
sky areas differ from the semi-analytic ones by up to $\ftGwShippedAreaPct\%$: it conjugates the
LAL polarizations into the $e^{+i\omega t}$ convention but applies the arrival-time delay with
the sign of the $e^{-i\omega t}$ convention, so the antenna-pattern and delay derivatives interfere
with the wrong relative sign.  Timing-only quantities are unaffected.  With the delay sign made
consistent, all marginal widths and areas agree to $\ftGwMargAgreePct\%$
(Table~\ref{tab:fisher_network}); its phase parameter is twice the reference phase used here.

\section{The extrinsic sampler and seeding menu}
\label{ap:sampler_menu}
Section~\ref{sec:jax_ile:sampling} defines the posterior over the extrinsic coordinates that
remain after direct marginalization.  Here we make explicit how the choice of marginalized
coordinates, sampler, and evidence estimator fit together.  Let
\(\bm\eta_{\mathsf M}\) denote the coordinates integrated by a selected marginalization scheme
\(\mathsf M\), and let \(\bm\theta_{\mathsf M}\) contain the remaining sampled coordinates.  The
sampler then sees
\begin{align}
\mathcal L_{\mathsf M}(\bm\lambda,\bm\theta_{\mathsf M})
 &=\int d\bm\eta_{\mathsf M}\,
 p(\bm\eta_{\mathsf M}\mid\bm\theta_{\mathsf M})
 \mathcal L(\bm\lambda,\bm\theta_{\mathsf M},\bm\eta_{\mathsf M}),
 \label{eq:sampler-menu:marginal-likelihood}\\
\pi_{\beta,\mathsf M}(\bm\theta_{\mathsf M})
 &=\frac{\mathcal L_{\mathsf M}^{\,\beta}
 p(\bm\theta_{\mathsf M})}{\mathcal Z_{\beta,\mathsf M}},
 \qquad
\mathcal Z_{\beta,\mathsf M}
 =\int d\bm\theta_{\mathsf M}\,
 \mathcal L_{\mathsf M}^{\,\beta}p(\bm\theta_{\mathsf M}).
 \label{eq:sampler-menu:tempered-target}
\end{align}
Here \(p(\bm\theta_{\mathsf M})\) is the corresponding marginal prior and \(\beta\) is the
inverse temperature.  The posterior and marginal likelihood required by RIFT are obtained at
\(\beta=1\).  Thus a
change of marginalization scheme changes the dimension and geometry of the sampled target, while
a change of sampler leaves Eqs.~\eqref{eq:sampler-menu:marginal-likelihood} and
\eqref{eq:sampler-menu:tempered-target} unchanged.
\begin{table*}
\centering
\caption{\label{tab:sampler_menu}\textbf{Sampler and estimator combinations available to the
differentiable extrinsic driver.}  The marginalized column lists the mode-dependent angular and
distance integrals; the time treatment is configured independently as described in
Appendix~\ref{ap:quadrature}.  ``IS'' denotes Eq.~\eqref{eq:sampler-menu:is}, and ``mixture IS''
uses one proposal component per recovered mode.  The SMC cloud is a modifier of the
phase- and polarization-marginalized flow modes rather than a distinct coordinate map.}
\begin{ruledtabular}
\begin{tabular}{llll}
mode & marginalized & sampling or proposal family & normalization \\
\hline
prior Monte Carlo      & optional distance & extrinsic prior                         & prior mean \\
adaptive Gaussian IS   & optional distance & iteratively fitted Gaussian             & IS \\
maximization           & none              & projected quasi-Newton search           & none \\
NUTS                   & distance          & one Hamiltonian chain                    & IS \\
multi-start NUTS       & distance          & one chain per recovered sky mode         & mixture IS \\
flow                   & distance          & normalizing flow with local moves        & IS \\
flow, phase-marg.      & distance, \(\phi_{\rm ref}\) & normalizing flow with local moves & IS or TI \\
flow, pol.-marg.       & distance, \(\psi\) & normalizing flow with local moves       & IS or TI \\
flow, phase/pol.-marg. & distance, \(\phi_{\rm ref},\psi\) & normalizing flow with local moves & IS or TI \\
Fisher-whitened NUTS   & distance, \(\phi_{\rm ref}\) & whitened chains by recovered mode & mixture IS \\
adaptive SMC cloud     & selected as above & tempered, resampled walker cloud         & SMC product or IS \\
\end{tabular}
\end{ruledtabular}
\end{table*}
Table~\ref{tab:sampler_menu} separates posterior exploration from normalization.  For prior Monte
Carlo, the proposal is \(q=p\), so the normalization is the prior mean of
\(\mathcal L_{\mathsf M}\).  The Gaussian, NUTS, flow, and cloud-fitted estimators instead draw
independently from a normalized proposal \(q\) and use
\begin{equation}
\widehat{\mathcal Z}_{\rm IS}
 =\frac{1}{N}\sum_{i=1}^{N} w_i,
\qquad
w_i=\frac{\mathcal L_{\mathsf M}(\bm\lambda,\bm\theta_i)
p(\bm\theta_i)}{q(\bm\theta_i)},
\qquad \bm\theta_i\sim q.
\label{eq:sampler-menu:is}
\end{equation}
A multi-start run uses
\(q=\sum_{m=1}^{K}a_m q_m\), where each \(q_m\) is fitted to one of the \(K\) recovered sky
modes and \(\sum_m a_m=1\).  Evaluating the full mixture density in every weight is essential:
treating the chains independently would leave their relative posterior masses undetermined.
Maximization is the exception in the table.  It returns
\(\max_{\bm\theta_{\mathsf M}}\mathcal L_{\mathsf M}\), not an integral, and therefore cannot
supply the marginal likelihood consumed by the intrinsic stage.
The adaptive cloud uses the sequence in Eq.~\eqref{eq:sampler-menu:tempered-target}.  If
\(W_i^{(k-1)}\) are normalized weights at \(\beta_{k-1}\), the incremental factors and
normalization update are
\begin{equation}
r_i^{(k)}=\mathcal L_{\mathsf M}(\bm\lambda,\bm\theta_i)^{
\beta_k-\beta_{k-1}},
\qquad
\frac{\widehat{\mathcal Z}_{\beta_k,\mathsf M}}
{\widehat{\mathcal Z}_{\beta_{k-1},\mathsf M}}
=\sum_i W_i^{(k-1)}r_i^{(k)} .
\label{eq:sampler-menu:smc-update}
\end{equation}
Multiplying these ratios from \(\beta_0=0\) to \(\beta_K=1\) gives the SMC normalization.  The
next rung is chosen from the effective sample size of the incremental weights; resampling and
local moves then repopulate the weighted region before the next update.  Once the terminal cloud
has adequate coverage, it can also define the defensive proposal used in
Eq.~\eqref{eq:sampler-menu:is}, as described in Section~\ref{sec:jax_ile:samplers}.  A tempered
flow uses the same family of targets but computes the normalization by thermodynamic integration
across the ladder.  These two ladder estimators are distinct even though both use
\(\beta\in[0,1]\).
Tempering also has a second, easily confused use.  In the production adaptive integrators,
tempering can broaden only the proposal used to place samples; the final importance weight still
has the form in Eq.~\eqref{eq:sampler-menu:is}.  In a differentiable tempered chain,
\(\beta<1\) instead changes the sampled target itself.  A draw from \(\pi_{\beta,\mathsf M}\)
must then be reweighted by \(r=\mathcal L_{\mathsf M}^{1-\beta}\) to represent the posterior.
For independent draws, the asymptotic retained fraction is
\begin{equation}
\frac{N_{\rm eff}}{N}
=\frac{\bigl\langle r\bigr\rangle_{\pi_\beta}^{2}}
{\bigl\langle r^2\bigr\rangle_{\pi_\beta}}
=\frac{\mathcal Z_{1,\mathsf M}^{\,2}}
{\mathcal Z_{\beta,\mathsf M}\mathcal Z_{2-\beta,\mathsf M}}.
\label{eq:sampler-menu:tempering-ess}
\end{equation}
In the local Gaussian limit with a broad prior and \(d\) sampled directions,
\(\mathcal Z_{a,\mathsf M}\propto a^{-d/2}\), and therefore
\begin{equation}
\frac{N_{\rm eff}}{N}=\bigl[\beta(2-\beta)\bigr]^{d/2}.
\label{eq:temper:cost}
\end{equation}
Equation~\eqref{eq:temper:cost} is a local-Gaussian estimate, not a general identity.  It shows
why a temperature chosen merely to broaden a proposal cannot be transferred to a chain that
samples the tempered target: terminal reweighting can discard most of the draws, with a penalty
that grows with the dimension left after marginalization.
Initialization determines which modes the subsequent adaptation can reach.  Each mode begins
with a pilot draw from the extrinsic prior.  High-likelihood pilot points provide initial chain
states, and the multi-start construction retains representatives of distinct sky modes rather
than only the global maximum.  Gradient ascent followed by Newton refinement can polish a seed;
the observed information at the polished point can then whiten a NUTS chain.  These operations
improve local adaptation but do not establish global coverage.  Coverage must instead be checked
from independent starts or from a cloud that retains all relevant modes.
A fitted flow can also initialize the next intrinsic point, but it is not a safe replacement for
that coverage check.  It must be treated as a proposal and corrected with the exact weight in
Eq.~\eqref{eq:sampler-menu:is}; a poorly matched proposal can otherwise contract the recovered
posterior or leave too few effective samples after correction.  The cloud-reuse construction in
Eq.~\eqref{eq:cloud_reuse} instead rebuilds a defensive proposal from corrected samples at every
intrinsic point.  It is not a selectable seeding mode in the differentiable driver.  More
generally, every warm start must either preserve support over the target or be combined with an
exact correction and an effective-sample-size check.

\clearpage

\end{document}